\documentclass[final]{elsart}
\journal{Journal of Geometry and Physics}

\usepackage{amssymb}
\usepackage{amsmath}
\usepackage[bookmarks,colorlinks=false]{hyperref}

\font\SYM=msbm12
\newcommand{\Real}{\mbox{\SYM R}}
\newcommand{\Complex}{\mbox{\SYM C}}

\newtheorem{proposition}{Proposition}

\newtheorem{theorem}{Theorem}

\newtheorem{corollary}{Corollary}
\newtheorem{remark}{Remark}

\newcommand{\DAl}{\raise -0.4mm \hbox{\Large$\Box$}}
\newcommand{\DAlhat}{{\raise -0.4mm \hbox{\Large$\Box$}}\hspace{-12pt}{\raise -1mm \hbox{\Large$\hat{\phantom{\Psi}}$}}}

\def\N{\hfill $\Box$}

\newcommand{\updn}[3]{#1^{#2}_{\phantom{#2}#3}}
\newcommand{\dnup}[3]{#1_{#2}^{\phantom{#2}#3}}

\begin{document}

\begin{frontmatter}

\title{Killing spinor initial data sets}

\author[author1]{Alfonso Garc\'{\i}a-Parrado G\'omez-Lobo}
\ead{algar@mai.liu.se}

\address[author1]{Matematiska institutionen, Link\"opings Universitet,\\
SE-581 83 Link\"oping, Sweden}

\author[author2]{Juan A. Valiente Kroon}
\ead{j.a.valiente-kroon@qmul.ac.uk}

\address[author2]{School of Mathematical Sciences, Queen Mary, University of London, \\
Mile End Road, London E1 4NS, UK.}




\begin{abstract}
  A 3+1 decomposition of the twistor and valence-2 Killing spinor equation is
  made using the space spinor formalism. Conditions on initial data
  sets for the Einstein vacuum equations are given so that their
  developments contain solutions to the twistor and/or Killing
  equations. These lead to the notions of twistor and Killing spinor
  initial data. These notions are used to obtain a characterisation of
  initial data sets whose development are of Petrov type N or D. 

\end{abstract}

\end{frontmatter}

\section{Introduction}
In this work we study the kind of conditions to be imposed on an
initial data set $(\mathcal{S},h_{ij},K_{ij})$ of the Einstein vacuum
field equations for its development $(\mathcal{M},g_{\mu\nu})$ to be
endowed with a \emph{Killing spinor}.  For an $r$-valence Killing
spinor, it will be understood a totally symmetric spinor
$\kappa_{(AB\cdots P)}=\kappa_{AB\cdots P}$, of valence $r$ such that
\[
\nabla_{Q'(Q}\kappa_{AB\cdots P)}=0,
\]   
although a more general definition exists with an arbitrary number of
symmetrised primed indices ---see \cite{PenRin86}. For reasons of
physical interest to be elaborated below, our analysis will be
concentrated on the cases with $r=1,2$, namely
\begin{subequations}
\begin{eqnarray}
&& \nabla_{A'(A}\kappa_{B)}=0, \label{twistor_eqn}\\
&& \nabla_{A'(A}\kappa_{BC)}=0. \label{Kspinor_eqn}
\end{eqnarray}
\end{subequations}
The interest in spinors satisfying equations (\ref{twistor_eqn}) or
(\ref{Kspinor_eqn}) stems from their potential use in the characterisation
of initial data sets yielding developments of a particular
Petrov type. An ultimate goal of this analysis is to yield a
characterisation of initial data sets for the Schwarzschild or Kerr
spacetimes.

Equation (\ref{twistor_eqn}) is usually called the \emph{twistor
  equation}. It is well known that its solvability imposes very strong
restrictions on the curvature (Weyl tensor) of the spacetime. One of these restrictions is
\begin{equation}
\Psi_{ABCQ}\kappa^Q=0, \label{typeN}
\end{equation}
where $\Psi_{ABCD}$ is the Weyl spinor ---see section
\ref{section:conventions} for more information on the conventions being used. The latter condition will 
occur if and only if the spacetime is of
Petrov type N ---see \cite{PenRin86}.  Therefore if we are able to find conditions on a
vacuum initial data set ensuring the existence of a spinor $\kappa_A$
fulfilling the twistor equation on at least an open subset of the
development, then these conditions will also guarantee that such a
subset is of Petrov type N.  These explicit conditions are presented
in theorem \ref{typeN:data}.

Equation (\ref{Kspinor_eqn}) has more physical relevance, as it can be
shown that all Petrov type D vacuum solutions to the Einstein field
equations ---and in particular the Schwarzschild and Kerr
spacetimes--- possess a valence-2 Killing spinor \cite{WalPen70} 
---see proposition \ref{theorem:rigidity} in the appendix.
In any case, if the spacetime has a valence-2 Killing spinor then it has
to be algebraically special and must satisfy the integrability condition
\begin{displaymath}
\dnup{\Psi}{(ABC}{Q} \kappa_{D)Q}=0.
\end{displaymath}
Further, if the valence-2 Killing spinor is non-null ---i.e. its
principal spinors are not repeated--- then it has to be of Petrov type
D \cite{Jef84}. If the valence-2 Killing spinor is null, then it can
be shown that the Weyl tensor has to be of Petrov type N ---see again
proposition \ref{theorem:rigidity} in the appendix for a proof of this
statement. Hence if we could find conditions ensuring that a
non-null valence-2 Killing spinor exists in at least a neighbourhood
of the initial data set then we could conclude that in such a
neighbourhood the spacetime is of type D.  This idea is the core of
theorem \ref{theorem:characterisationtypeD}.

Another interesting feature about vacuum spacetimes admitting a spinor
$\kappa_{AB}$ solving (\ref{Kspinor_eqn}) is that the vector $\xi^\mu$, defined in terms of its associated spinor $\xi_{AA'}$ by
\begin{equation}
\xi_{AA'}\equiv\nabla^Q_{\phantom{Q}A'} \kappa_{AQ} \label{Kvector}
\end{equation} 
is a ---possibly complex--- \emph{Killing vector}. Thus, one has
\begin{displaymath}
\nabla_{AA'}\xi_{BB'}+ \nabla_{BB'} \xi_{AA'}=0.
\end{displaymath}

This paper is organised as follows: basic conventions followed in this
paper are given in section \ref{section:conventions}. In section
\ref{sec:space_spinor}, we review the {\em space spinor} formalism,
which is essential to find the orthogonal decomposition of spinor
expressions.  Section \ref{section:symhyp} is provides a result on
wave equations which will be used extensively. Section
\ref{section:Kvector} is devoted to the study of conditions ---Killing
initial data, KID--- which ensure the existence of Killing vectors in
the development of a vacuum initial data set.  Although these
conditions are well-known, they are usually derived and presented in a
tensor form which is not suitable for this work. We carry out this
derivation in spinor form, so that it also illustrates procedures to
be used later. In section \ref{section:twistor_data}, we find the
necessary and sufficient conditions which a vacuum initial data set
must fulfil for the existence of a twistor in a neighbourhood of the
data development.  Similarly, in section
\ref{section:killing_spinor_data} conditions are derived for the
existence of a valence-2 Killing spinor in the data development. The
main results of the paper are contained in theorem \ref{typeN:data},
where it is shown when the development of a vacuum initial data set
has a Petrov type N region, and in theorem
\ref{theorem:characterisationtypeD} which provides a similar result
for Petrov type D.

In order to prove some of the results presented in this paper, long
computations with spinors are needed. These calculations have been
performed with the computer algebra system {\em xAct} \cite{xAct},
which is a suite of MATHEMATICA packages tailored for tensor calculus.
In particular, it can neatly handle all the essential rules of the
spinor calculus.
 
It should be noted that several of our results could be reformulated
in the language of Killing tensors and Killing-Yano tensors ---see
e.g. \cite{PenRin86}. As it is often the case, we expect the tensorial
expressions to be much more complicated that their spinorial
counterparts. However, for some particular applications it may prove
useful to have a tensorial expression.

\section{Conventions} \label{section:conventions} Let $({\mathcal M},
g_{\mu\nu})$ be a smooth Lorentzian manifold.  Due to the systematic
use of spinors in our discussion, the spacetime metric will be taken
to have signature $(+,-,-,-)$. We shall follow the conventions of
\cite{PenRin84,PenRin86} for spinor calculus.  Abstract indices are
used throughout with Greek letters denoting spacetime abstract indices
and capital Latin letters ---with or without dash--- denoting abstract
spin indices. The Riemann, Ricci and Weyl tensors will be denoted,
respectively, by $R_{\mu\nu\alpha\beta}$, $R_{\mu\nu}$,
$C_{\mu\nu\alpha\beta}$. The volume element is
$\eta_{\mu\nu\alpha\beta}$. We define the Hodge dual in the standard
fashion.  It will be assumed that $(\mathcal{M},g_{\mu\nu})$ is a
globally hyperbolic spacetime. This implies the existence of a smooth
time function $t:{\mathcal M}\rightarrow\mathbb R$ foliating $\mathcal
M$ ---see \cite{sanchez}.  Let $\mathcal{S}_s=\{p\in \mathcal{M} \;
|\; t(p)=s, s\in{\mathbb R} \}$ denote a leaf of the foliation of
$\mathcal{M}$ induced by $t$. Then ${\mathcal S}_s$ is an embedded
spacelike hypersurface on $\mathcal M$, for all $s\in{\mathbb R}$.
The timelike 1-form $\mbox{d} t$ is orthogonal to the leaves of the
foliation. We can construct another 1-form $\tau_\mu$ such that
$\tau_{\mu}\tau^{\mu}=2$ ---the reasons for this normalisation will
become apparent later.  From $\tau_\mu$ and $g_{\mu\nu}$ we introduce
the tensor $h_{\mu\nu}$ by means of the definition
\begin{displaymath}
 h_{\mu\nu}\equiv-\frac{1}{2}\tau_\mu \tau_\nu+g_{\mu\nu}
\end{displaymath}  
The tensor $h_{\mu\nu}$ plays the role of the intrinsic metric (first
fundamental form) of $\mathcal{S}_s$, for all $s\in{\mathbb R}$. Note
that this intrinsic metric is negative definite, due to the signature
convention chosen for $g_{\mu\nu}$. Using $\tau_\mu$ and $h_{\mu\nu}$
one can define all the mathematical objects used in the standard 3+1
decomposition. Our notation for the spatial derivative is $D_\mu$ and
our conventions for the second fundamental form $K_{\mu\nu}$ and the
acceleration $A^{\mu}$ are
\begin{eqnarray*}
&&K_{\mu\nu}=-h_\mu^{\ \alpha}h_{\nu}^{\ \beta}\nabla_\alpha\tau_\beta, \\ 
&& A^{\mu}=-\frac{1}{2}n^{\alpha}\nabla_{\alpha}n^{\mu}.
\end{eqnarray*} 
Any covariant spatial tensor $T$ corresponds to a unique tensor field
defined on any of the leaves ${\mathcal S}_s$ which is obtained by
means of the pull-back ${\mathfrak i}^*$ where ${\mathfrak
  i}:{\mathcal S}_s\rightarrow {\mathcal M}$ is the inclusion
embedding. The tensor field ${\mathfrak i}^*T$ is an element of the
tensor algebra constructed by taking ${\mathcal S}_s$ as the base
manifold. Latin characters will be used as the abstract indices
of this tensor algebra and therefore if $T_{\mu}$ is spatial then
$T_j$ will denote its pull-back under ${\mathfrak i}$.

\section{The space spinor formalism}
\label{sec:space_spinor}
It is of interest to find a spinor formulation of the standard 3+1
decomposition.  It can be seen that one can construct
spinors which transform under the group $SU(2)$. These spinors play
the role of the ``spatial elements'' and shall be called space spinors
or $SU(2)$ spinors. Accounts of the definition and main properties of
space spinors can be found in \cite{Som80,Ash91,Fra98a}. Here we will
review the basics of this formalism as it is an essential tool in this
work.

Let $\tau_{AA'}$ be the spinor equivalent of $\tau_{\mu}$ 
\footnote{Here and in the sequel, the spinor associated
  to a tensor, say $\dnup{\Phi}{\mu}{\nu}$, is  the
  spinor $\dnup{\Phi}{AA'}{BB'}= \updn{\sigma}{\mu}{AA'}
  \dnup{\sigma}{\nu}{BB'}\dnup{\Phi}{\mu}{\nu}$, where
  $\dnup{\sigma}{\mu}{AA'}$ denotes the {\em soldering form} and
  $g_{\mu\nu}=\epsilon_{AB} \epsilon_{A'B'}\dnup{\sigma}{\mu}{AA'}
  \dnup{\sigma}{\nu}{BB'}$.}. 
Clearly, $\tau_{AA'}\tau^{AA'}=2$ and 
$$
\tau^{A}_{\phantom{A}A'}\tau^{BA'}=\epsilon^{AB}.
$$
Using this spinor we may define the {\em spatial soldering form} as  
\begin{displaymath}
\dnup{\sigma}{\mu}{AB}\equiv
\sigma_{\mu\phantom{(A}A'}^{\phantom{\mu}(A}\tau^{B)A'}.
\end{displaymath} 
The spatial soldering form $\sigma_\mu^{\ AB}$ should not be confused
with the soldering form $\sigma_\mu^{\ AA'}$. They are in fact
different objects despite having the same kernel letter.  The
algebraic properties of the spatial soldering form are
\begin{equation}
\updn{\sigma}{\mu}{AB}\dnup{\sigma}{\nu}{AB}=h^{\mu}_{\phantom{\mu}\nu},\quad
\updn{\sigma}{\mu}{AB}\sigma_{\mu CD}=
\frac{1}{2}(\epsilon_{AC}\epsilon_{BD}+\epsilon_{AD}\epsilon_{BC}),\quad 
\tau_\mu\updn{\sigma}{\mu}{AB}=0.
\label{spatial_soldering_properties}
\end{equation} 
Inspecting these algebraic properties one can see that
$\updn{\sigma}{\mu}{AB}$ can be used to define a spin structure
associated to the metric $h_{\mu\nu}$.  Spinors belonging to such spin
structure are the space spinors mentioned above and it is possible to
relate spatial tensors to space spinors and viceversa by means of the
spatial soldering form. As an example of this let $\dnup{W}{i}{j}$ be
a spatial tensor. Its space-spinor counterpart is given by
$\dnup{W}{AB}{CD}= \updn{\sigma}{i}{AB} \dnup{\sigma}{j}{CD}
\dnup{W}{i}{j}$. Another important example is the spatial metric
$h_{ij}$ whose space spinor equivalent can be obtained by using the
second and third expressions of (\ref{spatial_soldering_properties}):
\begin{displaymath}
 h_{ABCD}=\frac{1}{2}(\epsilon_{AC}\epsilon_{BD}+\epsilon_{AD}\epsilon_{BC}).
\end{displaymath} 
In order to avoid complicating the notation, the same Kernel letter
will be used to denote a spatial tensor and its space spinor
equivalent; the nature of the relevant object will be indicated in the
text or in most cases it will be clear from the context.

An important issue is to find when a given $SL(2,\Complex)$ spinor
arises from a spatial tensor ---and thus is a $SU(2)$ spinor.
In order to answer this question we need to introduce the Hermitian
conjugation operation. Given any spinor $Z_{A_1\cdots A_n}$ we define
its Hermitian conjugate as
\begin{displaymath}
\widehat{Z}_{A_1\cdots A_n}\equiv 
\tau_{A_1}^{\phantom{A_1}A'_1}\cdots\tau_{A_n}^{\phantom{A_n}A'_n}
\overline{Z}_{A'_1\cdots A'_n}, 
\end{displaymath}
where the overline denotes the standard complex conjugation.  Any even
rank spinor $W_{A_1B_1\cdots A_nB_n}$ stems from a real spatial tensor
if and only if the following two conditions hold:
\begin{subequations}
\begin{eqnarray}
&& \widehat{W}_{A_1B_1\cdots A_nB_n}
=(-1)^{n}W_{A_1B_1\cdots A_nB_n}, \label{space_spinor_condition}\\ 
&& W_{A_1B_1\cdots A_nB_n}=W_{(A_1B_1)\cdots (A_nB_n)}.
\end{eqnarray}
\end{subequations} 
Any spinor of even rank fulfilling condition
(\ref{space_spinor_condition}) will be said to be {\em real}.
Therefore we deduce that there exists an equivalence between real and
space spinors. If for a spinor $W_{A_1B_1\cdots A_nB_n}=W_{(A_1B_1)\cdots (A_nB_n)}$ the condition
\begin{displaymath}
\widehat{W}_{A_1B_1\cdots A_nB_n} =(-1)^{n+1}W_{A_1B_1\cdots A_nB_n},
\end{displaymath}
holds, then the spinor is said to be \emph{imaginary}.

As in the case of spatial tensors, space spinors can also be regarded
as {\em intrinsic objects} in any of the Riemannian manifolds
${\mathcal S}_s$. This means that there is an isomorphism between the
subalgebra of space spinors and a tensor algebra constructed from a
vector bundle arising from a spin structure on $({\mathcal S}_s,
h_{ij})$. In order to avoid excessive notation we are not going to
define new abstract indices for this vector bundle and instead we will
simply add a tilde over the kernel letter of a space spinor whenever
we wish to consider it as an element of the spin bundle constructed
over $({\mathcal S}_s, h_{ij})$.

\subsection{Spatial spin covariant derivatives}
\label{covariant_derivative}
The spinorial covariant derivative can be decomposed as follows
\begin{equation}
\nabla_{AA'}= \frac{1}{2} \tau_{AA'} \nabla -\dnup{\tau}{A'}{C} \nabla_{AC},
\label{spin-decomposition}
\end{equation}
where
\begin{eqnarray*}
&& \nabla\equiv\tau_{AA'} \nabla^{AA'}, \\
&& \nabla_{AB}\equiv\updn{\tau}{A'}{(A}\nabla_{B)A'}=\updn{\sigma}{\mu}{AB}\nabla_\mu.
\end{eqnarray*}
The operator $\nabla_{AB}$ is usually referred to as the \emph{Sen connection}.
Next we introduce the spinors 
\begin{eqnarray*}
&& K_{AB}\equiv\dnup{\tau}{B}{A'}\nabla \tau_{AA'}, \\
&& K_{ABCD}\equiv\dnup{\tau}{D}{C'}\nabla_{AB}\tau_{CC'}.
\end{eqnarray*}
These spinors satisfy the following algebraic properties
\begin{subequations}
\begin{eqnarray}
&& K_{(AB)}=K_{AB},\ \widehat{K}_{AB}=-K_{AB},\quad   K_{(AB)(CD)}=K_{ABCD}, \label{k-properties1}\\
&& \widehat{K}_{ABCD}=K_{ABCD}, \quad  K_{FD}=-2 A_{\mu}\updn{\sigma}{\mu}{FD},\label{k-properties2}\\ 
&& K_{ABCD}=-\updn{\sigma}{\mu}{AB}\updn{\sigma}{\nu}{CD}K_{\mu\nu}. 
\label{k-properties3}
\end{eqnarray}
\end{subequations} 
The spinor $K_{AB}$ is called the \emph{acceleration} spinor while
$K_{ABCD}$ corresponds to the second fundamental form of the leaves.
Formulae (\ref{spin-decomposition}) and
(\ref{k-properties1})-(\ref{k-properties3}) hold regardless to whether
$\tau_\mu$ is hypersurface forming or not ---if $\tau_{\mu}$ is not
hypersurface forming then $K_{\mu\nu}$ fails to be symmetric. In the
case of $\tau_\mu$ being integrable then we get the extra symmetry
\begin{equation}
K_{ABCD}=K_{CDAB},
\label{extra-symmetry}
\end{equation}
which is a straightforward consequence of $K_{\mu\nu}=K_{\nu\mu}$. As
we are working with an hypersurface forming $\tau_\mu$ we will take
for granted condition (\ref{extra-symmetry}) in what is to follow.

Many of our arguments will make use of a foliation for which
$K_{AB}=0$. From (\ref{k-properties1})-(\ref{k-properties3}) it  
follows directly that $A_{\mu}=0$, that is, $\tau_{AA'}$ is geodesic. 
Such a foliation can always be constructed in, at least, 
a neighbourhood of any spacelike
hypersurface. 

The operator $\nabla_{AB}$ corresponds to the operator
$\dnup{h}{\mu}{\nu}\nabla_\nu$ which is not intrinsic to the
hypersurfaces $\mathcal{S}_s$ ---the action of $\nabla_{AB}$ on a
space spinor does not result in a space spinor. In order to obtain a
differential operator which maps space spinors into space spinors let
us start by defining 
\begin{eqnarray*}
&& D \pi_A\equiv \nabla \pi_A -\frac{1}{2} \dnup{K}{A}{B}\pi_B,\quad  
D\overline{\pi}_{A'}\equiv\nabla\overline{\pi}_{A'}-\frac{1}{2}\dnup{\overline{K}}{A'}{B'}
\overline{\pi}_{B'},\\
&& D_{AB} \pi_C\equiv\nabla_{AB}\pi_C -\frac{1}{2} \dnup{K}{ABC}{D}\pi_D, \\ 
&& D_{AB}\overline{\pi}_{C'}\equiv\nabla_{AB}\overline{\pi}_{C'}+
\frac{1}{2}K_{ABCD}\tau^{CA'}\tau^{D}_{\phantom{D}C'}\overline{\pi}_{A'}.
\end{eqnarray*}
The operators $D$, $D_{AB}$ are extended to the full spinor algebra by
requiring them to satisfy the Leibnitz rule. Important properties
of these operators are
\begin{displaymath}
 D\tau^{A}_{\phantom{A}B'}=0,\quad D_{AB}\tau^{C}_{\phantom{A}C'}=0,\quad 
D\epsilon_{AB}=0,\quad   D_{AB}\epsilon_{CD}=0. 
\end{displaymath}
Using these properties and equation (\ref{space_spinor_condition}) it is now a simple matter to check that the action of $D_{AB}$ on a space spinor is again a space spinor.  Therefore whenever
$\tau_{AA'}$ is surface forming, the operator $D_{AB}$ can be regarded as the spinorial
counterpart of the spatial derivative $D_\mu$. 

In the sequel, the following commutators will be used
\begin{subequations}
\begin{eqnarray}
&&[\nabla, \nabla_{AB}] \alpha_C = -\frac{1}{2} K_{AB} \nabla \alpha_C -\dnup{K}{(A}{Q} \nabla_{B)Q} \alpha_C -K_{ABPQ}\nabla^{PQ}\alpha_C+\nonumber\\
&&\hspace{3.5cm}+\DAlhat_{AB} \alpha_C-\DAl_{AB} \alpha_C ,\label{commutator1}\\
&& [\nabla_{AB},\nabla_{CD}] \alpha_E= \frac{1}{4}\left( \epsilon_{AC} \DAl_{BD} + \epsilon_{AD} \DAl_{BC} + \epsilon_{BC} \DAl_{AD} + \epsilon_{BD}\DAl_{AC} \right)\alpha_E\nonumber\\
&&\hspace{3.5cm}+\frac{1}{4}\left( \epsilon_{AC} {\DAlhat}_{BD} + \epsilon_{AD}{\DAlhat}_{BC} + \epsilon_{BC} {\DAlhat}_{AD} + \epsilon_{BD}{\DAlhat}_{AC} \right) \alpha_E\nonumber \\
&&\hspace{3.5cm}-\frac{1}{2} \left( K_{ABCQ}\dnup{\nabla}{D}{Q} + K_{ABDQ} \dnup{\nabla}{C}{Q}+K_{CDAQ} \dnup{\nabla}{B}{Q} + K_{CDBQ} \dnup{\nabla}{A}{Q} \right)\alpha_E, \nonumber \\
&&\label{commutator2}
\end{eqnarray}
\end{subequations}
where $\DAl_{AB}=\nabla_{C'(A} \dnup{\nabla}{B)}{C'}$, and
$\DAlhat_{AB}= \dnup{\tau}{A}{A'} \dnup{\tau}{B}{B'} \DAl_{A'B'}$.
The action of these operators on any spinor $\alpha_C$ is
\begin{equation}
\DAl_{AB}\alpha_C= \Psi_{ABCD} \alpha^D+\frac{\Lambda}{2}\epsilon_{C(A}\alpha_{B)}, 
\quad \DAlhat_{AB} \alpha_C =-\tau_A^{\phantom{A}A'}\tau_{B}^{\phantom{B}B'}\Phi_{FCA'B'}\alpha^F,
\label{box-relations}
\end{equation}
with $\Psi_{ABCD}$ the Weyl spinor, $\Lambda$ the scalar curvature and $\Phi_{ABA'B'}$ the Ricci spinor.


From (\ref{commutator1})-(\ref{commutator2}) we can obtain similar commutation relations for the operators $D$ and $D_{AB}$. These are
\begin{subequations}
\begin{eqnarray}
&&[D,D_{AB}]\alpha_C=-
K_{ABDF} D^{DF}\alpha_{C}-\frac{1}{2} K_{AB} D\alpha_{C}+\DAl_{AB}\alpha_{C}+\DAlhat_{AB}\alpha_{C}\nonumber\\
&&\hspace{3.5cm}+\frac{\alpha^D}{4}\left(K_{AB}K_{CD}-K_{DF}K_{ABC}^{\phantom{ABC}F}
-K_{CF}K_{ABD}^{\phantom{ABD}F}\right.\nonumber\\
&&\hspace{3.5cm}\left.+2 K_{ABFH}
K_{CD}^{\phantom{CD}FH} -2 D_{AB}K_{CD}+2 D K_{ABCD}\right),\label{space_comm1}\\
&& [D_{AB},D_{CD}]\alpha_F=\frac{\alpha_H}{4}\left(-K_{AB}^{\phantom{HL}HL}K_{CDFL}
+K_{CD}^{\phantom{HL}HL}K_{ABFL}+2D_{CD}K_{ABF}^{\phantom{ABF}H}-
2D_{AB}K_{CDF}^{\phantom{CDF}H}\right) \nonumber\\
&&\hspace{3.5cm}+\frac{1}{4}\left(\epsilon_{AC}\DAlhat_{BD}+\epsilon_{AD}\DAlhat_{BC}+
\epsilon_{BC}\DAlhat_{AD}+\epsilon_{BD}\DAlhat_{AC}\right)\alpha_F\nonumber\\
&&\hspace{3.5cm}+\frac{1}{4}\left(\epsilon_{AC}\DAl_{BD}+\epsilon_{AD}\DAl_{BC}+
\epsilon_{BC}\DAl_{AD}+\epsilon_{BD}\DAl_{AC}\right)\alpha_F.\label{space_comm2}
\end{eqnarray}
 
\end{subequations}
The commutation relations (\ref{commutator1})-(\ref{commutator2}) and
(\ref{space_comm1})-(\ref{space_comm2}) can be generalised if we let
the commutators act on spinors of higher rank. For instance if we
choose the spinor $\tau_{AA'}$ then we obtain the identities
\begin{subequations}
\begin{eqnarray}
&& D K_{AFCD}=2 (E_{ACDF}+\widehat{\Phi}_{ACDF})-
\frac{1}{2}K_{AF}K_{CD}-K_{AF}^{\phantom{AF}BH}K_{CDBH}\nonumber\\
&& \hspace{3.5cm}-2\Lambda(\epsilon_{AD}\epsilon_{CF}+\epsilon_{AC}\epsilon_{DF})+
D_{CD}K_{AF},\label{evolutioneqn:K_ABCD}\\
&& B_{ABCD}= -\mbox{i}\ \updn{D}{Q}{(A}K_{|BC|D)Q}, \label{B_ABCD}
\end{eqnarray}
\end{subequations}
where we have introduced the 
\emph{electric} and \emph{magnetic} parts of the Weyl spinor
$\Psi_{ABCD}$ which are given, respectively, by
\begin{displaymath}
E_{ABCD}=\frac{1}{2}(\hat{\Psi}_{ABCD}+\Psi_{ABCD}), \quad B_{ABCD}=\frac{\mbox{i}}{2}(\hat{\Psi}_{ABCD}-\Psi_{ABCD}).
\end{displaymath}
Note that both $E_{ABCD}$ and $B_{ABCD}$ are totally symmetric and real. 
They are related to the electric and magnetic parts of the Weyl tensor
by the relations
\begin{displaymath}
 \dnup{\sigma}{\mu}{AB}\dnup{\sigma}{\nu}{CD}E_{ABCD}=E_{\mu\nu},\quad 
 \dnup{\sigma}{\mu}{AB}\dnup{\sigma}{\nu}{CD}B_{ABCD}=B_{\mu\nu},
\end{displaymath}
with the definitions
\begin{displaymath}
 E_{\mu\nu}\equiv\frac{1}{2}\tau^\alpha\tau^\beta C_{\mu\alpha\nu\beta},\quad 
 B_{\mu\nu}\equiv\frac{1}{2}\tau^\alpha\tau^\beta C^*_{\mu\alpha\nu\beta},
\end{displaymath}
where
$C^*_{\mu\alpha\nu\beta}=\frac{1}{2}\eta_{\nu\beta\lambda\pi}C_{\mu\alpha}^{\phantom{\mu\alpha}\lambda\pi}$.
It is a property of the space-spinorial formalism that symmetric
trace-free spatial tensors are associated to totally symmetric
space-spinors.  

As discussed in \cite{Val05b,GarVal07} the values of $E_{\mu\nu}$ and
$B_{\mu\nu}$ on the initial hypersurface $\mathcal{S}$ can be
calculated entirely from the values of $h_{ij}$ and $K_{ij}$. This
is also true for their spinorial counterparts.  In the case of the magnetic part this assertion is obvious 
from (\ref{B_ABCD}). To obtain the corresponding expression of the electric part 
we start from the identity
\begin{equation}
(D_{CD}D_{FH}-D_{FH}D_{CD})\widetilde{\zeta}_{AB}=\widetilde{r}_{LPABCDFH}
\widetilde{\zeta}^{LP},
\label{dd_commutation}
\end{equation}
where $\widetilde{\zeta}_{AB}$ is any symmetric spinor and
$\widetilde{r}_{LPABCDFH}$ is a spinor which represents the spatial
Riemann tensor $\widetilde{r}_{ijkl}$. The explicit expression for
$\widetilde{r}_{LPABCDFH}$ can be obtained from the generalisation of
(\ref{space_comm2}) to $\widetilde{\zeta}_{AB}$ and shall be omitted
as it as it is somewhat long. As happens with the spinor representing
the spacetime Riemann tensor, the spinor $\widetilde{r}_{LPABCDFH}$
can be decomposed in irreducible parts which are obtained by taking
suitable traces. For instance, the totally symmetric part
$\widetilde{r}_{BCFL}$ is defined by
$$
\widetilde{r}_{BCFL}\equiv
-\widetilde{r}_{(L\phantom{A}|A|BC|\phantom{D}F)D}^{\phantom{(L}A\phantom{|A|BC|}D},
$$
from which using the expression for $\widetilde{r}_{ABCDFHLP}$ deduced
from (\ref{dd_commutation}) we get the desired expression for
$\widetilde{E}_{ABCD}$:
\begin{eqnarray*}
&&\widetilde{E}_{ABCD}= -\widetilde{r}_{ABCD} -\frac{1}{6}\widetilde{\Omega}_{ABCD}\widetilde{K} + \frac{1}{2}\dnup{\widetilde{\Omega}}{(AB}{PQ}\widetilde{\Omega}_{CD)PQ},\quad 
\widetilde{K}\equiv \widetilde{K}^{PQ}_{\phantom{PQ}PQ},\quad\\
&&\widetilde{\Omega}_{ABCD}\equiv\widetilde{K}_{(ABCD)}
\end{eqnarray*}

For completeness we include also the expressions for the Hamiltonian
and momentum constraints in the space spinor formalism. These are
\begin{eqnarray*}
&&\widetilde{r} +\frac{1}{6}\widetilde{K}^2-\frac{1}{4}\widetilde{\Omega}_{ABCD}
\widetilde{\Omega}^{ABCD}=0,\quad 
\widetilde{r}\equiv \widetilde{r}^{AC\phantom{C}D\phantom{A}L}_{\phantom{AC}C\phantom{D}A\phantom{L}DL},\\
&& D^{PQ}\widetilde{K}_{PQAB}-\frac{1}{2}D_{AB}\widetilde{K}=0.
\end{eqnarray*}
The Hamiltonian constraint is obtained by taking the suitable traces in the 
formula for $\widetilde{r}_{ABCDFHLP}$ while the momentum constraint is the nonvanishing trace of (\ref{B_ABCD}).

\section{Homogeneous second order hyperbolic systems}
\label{section:symhyp} 
In this work we will make use of the following
result which can be consulted for instance in \cite{Taylor96} ---p.
378, Proposition 3.2.

\begin{theorem}
  Let $g_{\mu\nu}$ be a Lorentzian metric on a smooth manifold
  $\mathcal M$ and define the differential operator $\DAl\equiv
  g^{\mu\nu}\nabla_\mu\nabla_\nu$ (D'Alembertian).  Consider the
  second order hyperbolic system
\begin{equation}
\DAl X + H(X,\nabla_{\mu}X) =0,
\label{hyperbolic_system}
\end{equation}
where $X=(X_0,\ldots,X_m)$ is a set of scalar functions on $\mathcal
M$ and $H(X,\nabla_\mu X)$ is a smooth linear, homogeneous function of
the components of $X$ and their first covariant derivatives
$\nabla_\mu X$. Let ${\mathcal S}\subset{\mathcal M}$ be a spacelike
hypersurface with respect to $g_{\mu\nu}$ and assume that
\begin{displaymath}
X|_\mathcal{S}=f, \quad \nabla_{AA'}X|_\mathcal{S}=g,
\end{displaymath}
where $f$, $g$ are smooth on $\mathcal{S}$.
Then there exists a unique smooth solution of (\ref{hyperbolic_system}) in a neighbourhood of 
$\mathcal S$. In particular if $f=0,\ g=0$ then such unique solution is given by $X=0$. 
\label{hyperbolic_propagation}
\end{theorem}

\section{Killing initial data} \label{section:Kvector} 
In this
section, we study the conditions under which the development of a
vacuum initial data set will contain Killing vectors. The contents of
this section follow the ideas and methods of \cite{Col77,Mon75,Chr91b}
but in our case the results are cast in the spinorial language and the
techniques used in the proofs will be needed later on. From this point
on, $({\mathcal M},g_{\mu\nu})$ will be assumed to be a vacuum
spacetime.

\begin{proposition} \label{proposition:Killing:propagation}
Let $\eta_{AA'}$ be a spinorial field on $\mathcal{M}$, $\mathcal{S}_0\subset\mathcal{M}$ 
a Cauchy spacelike hypersurface and suppose that the following conditions hold 
\begin{subequations}
\begin{eqnarray}
&&(\nabla_{AA'} \eta_{BB'} + \nabla_{BB'}\eta_{AA'})|_{\mathcal{S}_0} =0,  \label{eta_kid_1}\\
&& \nabla_{CC'} (\nabla_{AA'}\eta_{BB'} + \nabla_{BB'}\eta_{AA'})|_{\mathcal{S}_0}=0. \label{eta_kid_2}
\end{eqnarray}
\end{subequations}
Assume further that 
\begin{displaymath}
\DAl\eta_{AA'}=0,
\end{displaymath}
 in an open set $\mathcal{W}$ which contains $\mathcal{S}_0$. 
Then, there exists an open set ${\mathcal U}\subset{\mathcal W}$ containing $\mathcal{S}_0$ such that the condition
\begin{displaymath}
\nabla_{AA'} \eta_{BB'} + \nabla_{BB'} \eta_{AA'} =0,
\end{displaymath}
holds on $\mathcal U$.
\label{killing_propagation}
\end{proposition}
\noindent
\textbf{Proof.}  
Define 
\begin{equation}
S_{AA'BB'} \equiv \nabla_{AA'}\eta_{BB'} + \nabla_{BB'}\eta_{AA'}.
\end{equation}
A lengthy, but straight-forward calculation renders
\begin{eqnarray}
&&\DAl S_{AA'BB'} = \nabla_{AA'} \DAl\eta_{BB'} + \nabla_{BB'} \DAl\eta_{AA'}+ \nonumber\\ 
&&\hspace{3.5cm}+2\dnup{\Psi}{AB}{PQ} S_{PA'QB'} + 2\dnup{\overline{\Psi}}{A'B'}{P'Q'} S_{AP'BQ'}, \label{Killing:propagation1}\\
&& \DAl\eta_{BB'}=\nabla_{AA'}S_{BB'}^{\phantom{BB'}AA'}
-\frac{1}{2}\nabla_{BB'}S^{AA'}_{\phantom{AA'}AA'}\label{Killing:propagation2},
\end{eqnarray}
which can be regarded as a system of partial differential equations in the variables 
$\eta_{AA'}$, $S_{AA'BB'}$. In the open set $\mathcal{W}$ the system
 (\ref{Killing:propagation1})-(\ref{Killing:propagation2}) becomes 
\begin{subequations}
\begin{eqnarray}
&&\DAl S_{AA'BB'} = 2 \dnup{\Psi}{AB}{PQ} S_{PA'QB'} + 
2\dnup{\overline{\Psi}}{A'B'}{P'Q'} S_{AP'BQ'},\label{Killing:propagation:0}\\
&&\nabla_{AA'}S_{BB'}^{\phantom{BB'}AA'}
-\frac{1}{2}\nabla_{BB'}S^{AA'}_{\phantom{AA'}AA'}=0.\label{Killing:propagation:1}
\end{eqnarray}
\end{subequations}
This is to be supplemented by the initial conditions
\begin{equation}
S_{AA'BB'}|_{\mathcal{S}_0}=0, \quad  \nabla_{CC'}S_{AA'BB'}|_{\mathcal{S}_0}=0,
\label{initial_conditions}
\end{equation}
implied by (\ref{eta_kid_1})-(\ref{eta_kid_2}). Theorem \ref{hyperbolic_propagation}
tells us that the initial value problem posed by 
(\ref{Killing:propagation:0}) and the conditions (\ref{initial_conditions}) has the unique solution $S_{AA'BB'}=0$ in a neighbourhood $\mathcal{U}$ containing ${\mathcal S}_0$. Trivially, this solution is consistent with (\ref{Killing:propagation:1})  and therefore the proof is complete. \N 

\begin{remark} 
\em It is important to point out that although the result given by 
proposition \ref{section:Kvector}
is local, a global version which guarantees the existence of a Killing
vector on the whole of $\mathcal{M}$ can be obtained if the spacetime
and $\eta_{AA'}|_{\mathcal{S}_0}$ are suitably smooth. One needs to
ensure that the \emph{Killing vector candidate} $\eta_{AA'}$
satisfying $\DAl\eta_{AA'}=0$ exists on the whole of $\mathcal{M}$. An
example of the conditions ensuring this can be found in \cite{Chr91b}.
\end{remark}

Proposition \ref{killing_propagation} enables us to determine under
which conditions a Killing vector will exist in a neighbourhood of
${\mathcal S}_0$. In our framework it is important to express this
result in terms of objects which are intrinsic to ${\mathcal S}_0$ and
the resulting conditions are called {\em Killing initial data}.  In
order to find the Killing initial data let us start by finding the
orthogonal decomposition of the Killing equation
\begin{equation}
 \nabla_{AA'}\xi_{BB'}+\nabla_{BB'}\xi_{AA'}=0,
\label{Killing_eqn}
\end{equation}
with respect to $\tau_{AA'}$.
First, we decompose $\xi_{AA'}$ by writting 
\begin{displaymath}
\xi_{AA'} =\frac{1}{2}\tau_{AA'}\xi -\updn{\tau}{Q}{A'} \xi_{AQ}.
\end{displaymath}
Contracting the Killing equation (\ref{Killing_eqn}) in all possible
manners with $\tau_{AA'}$ and symmetrising when necessary one obtains
the following expressions:
\begin{subequations}
\begin{eqnarray}
&& \nabla \xi - K_{PQ} \xi^{PQ}=0,  \label{kid_a} \\
&& \nabla \xi_{BC} + \nabla_{BC} \xi + \frac{1}{2} K_{BC} \xi - K_{P(B} \dnup{\xi}{C)}{P}- K_{BCPQ} \xi^{PQ}=0, \label{kid_b} \\
&&  \nabla_{(AB} \xi_{CD)} +\frac{1}{2} K_{(ABCD)} \xi -2 \dnup{K}{(ABC}{Q}\xi_{D)Q}=0. \label{kid_3}
\end{eqnarray}
\end{subequations} 
where $\nabla_{AA'}$ has been decomposed in terms of $\nabla_{AB}$ and
$\nabla$ ---see (\ref{spin-decomposition}).  Equations
(\ref{kid_b})-(\ref{kid_3}) are equivalent to
\begin{subequations}
\begin{eqnarray}
&& D\xi_{CF}=K_{CFAL}\xi^{AL}-\frac{1}{2}K_{CF}\xi-D_{CF}\xi, \label{kid_c1}\\
&& D_{AB}\xi_{CD}+D_{CD}\xi_{AB}+\xi K_{ABCD}=0. \label{kid_c2}
\end{eqnarray}
\end{subequations}
Condition (\ref{kid_c2}) is intrinsic to the leaves of
$\mathcal{S}_s$ and in particular to ${\mathcal S}_0 $. If we apply
the operator $D$ to it one obtains
\begin{equation}
DD_{AB}\xi_{CD}+DD_{CD}\xi_{AB}+ K_{ABCD}D\xi+\xi D K_{ABCD}=0.
\label{d_killing}
\end{equation}
To see that this last expression is indeed intrinsic to ${\mathcal
  S}_0$ we transform it by using the commutator (\ref{space_comm1})
together with equations (\ref{kid_a}), (\ref{kid_c1}) and
(\ref{evolutioneqn:K_ABCD}). This yields
\begin{eqnarray}
&& D_{DL} D_{CF} \xi +D_{CF} D_{DL}\xi+K_{DL}^{\phantom{DL}AB}D_{AB}\xi_{CF}+
K_{CF}^{\phantom{DL}AB}D_{AB}\xi_{DL} \nonumber\\
&&\hspace{2.5cm}+\xi(2E_{CDFL}-K_{CF}^{\phantom{CF}AB}K_{DLAB})-D_{DL}(\xi^{AB}K_{CFAB})\nonumber\\
&&\hspace{2.5cm}-D_{CF}(\xi^{AB}K_{DLAB})+4\mbox{\em i}\ \xi^{A}_{\phantom{A}(C}B_{DFL)A}=0,\nonumber\\
&&\label{prekid_2}
\end{eqnarray}
where a foliation has been chosen for which the acceleration vanishes 
---as we discussed earlier this can always be done in a neighbourhood
of ${\mathcal S}_0$.

The aim of previous calculations is to show  that equations (\ref{kid_c2}) and (\ref{prekid_2}) will hold on
${\mathcal S}_0$ whenever a Killing vector $\eta_{AA'}$ exists.  The
next proposition asserts that there is a converse to this property.

\begin{proposition}[\bf{Killing initial data}]
  \label{proposition:Killing:data} Let $(\mathcal{S},h_{ij},K_{ij})$
  be an initial data set for the vacuum Einstein field equations and
  let $({\mathcal M},g_{\mu\nu})$ be the data development. Assume
  further that on $\mathcal S$ there exist a scalar $\widetilde{\xi}$
  and a space spinor $\widetilde{\xi}_{AB}$ satisfying the conditions
  (Killing initial data)
\begin{subequations}
\begin{eqnarray}
&& D_{AB}\widetilde{\xi}_{CD}+D_{CD}\widetilde{\xi}_{AB}+\widetilde{\xi}
\widetilde{K}_{ABCD}=0, \label{kid_1}\\
&& D_{DL} D_{CF} \widetilde{\xi} +D_{CF}D_{DL}\widetilde{\xi}+\widetilde{K}_{DL}^{\phantom{DL}AB}D_{AB}\widetilde{\xi}_{CF}+
\widetilde{K}_{CF}^{\phantom{DL}AB}D_{AB}\widetilde{\xi}_{DL}\nonumber\\
&&\hspace{2cm}-\widetilde{\xi}(2\widetilde{E}_{CDFL}-
\widetilde{K}_{CF}^{\phantom{CF}AB}\widetilde{K}_{DLAB})-
D_{DL}(\widetilde{\xi}^{AB}\widetilde{K}_{CFAB})\nonumber\\
&&\hspace{2cm}-D_{CF}(\widetilde{\xi}^{AB}\widetilde{K}_{DLAB})\label{kid_2}+
4\mbox{\em i}\ \widetilde{\xi}^{A}_{\phantom{A}(C}\widetilde{B}_{DFL)A}=0.
\end{eqnarray}
\end{subequations}
Then there exists a spinorial field $\mathring{\xi}_{AA'}$  which is a Killing vector of $g_{\mu\nu}$ on a neighbourhood of $\mathcal{M}$.
\end{proposition}
\noindent
\textbf{Proof.}  Construct a foliation in the development $\mathcal M$
with leaves ${\mathcal S}_s$ such that $\mathcal{S}_0$ is identified
with $\mathcal{S}$ and $K_{AB}=0$ in at least a neighbourhood of
$\mathcal{S}_0$.  If $\tau_{AA'}$ is the normal to this foliation we
define a spacetime spinor $\mathring{\xi}_{AA'}$
\begin{equation}
\mathring{\xi}_{AA'}= \frac{1}{2}\mathring{\xi} \tau_{AA'} - \updn{\tau}{Q}{A'} \mathring{\xi}_{QA},
\quad \mathring{\xi}_{(AB)}=\mathring{\xi}_{AB}, \label{split}
\end{equation}
---the \emph{Killing vector candidate}--- by requiring that it satisfies
the wave equation
\begin{displaymath}
\DAl \mathring{\xi}_{AA'}=0, \label{wave_xi}
\end{displaymath}
supplemented with the initial conditions
\begin{subequations}
\begin{eqnarray}
&&\mathring{\xi}|_{{\mathcal S}_0}=\widetilde{\xi},\quad 
\mathring{\xi}_{AB}|_{{\mathcal S}_0}=\widetilde{\xi}_{AB},\label{propagation:rule_a} \\
&& 
D\mathring{\xi}|_{\mathcal{S}_0}=0, \quad D\mathring{\xi}_{AB}|_{\mathcal{S}_0} =-D_{AB}\widetilde{\xi}+\widetilde{K}_{ABPQ} \widetilde{\xi}^{PQ}. \label{propagation:rule_b}
\end{eqnarray}
\end{subequations}
According to theorem \ref{hyperbolic_propagation}
a solution for this initial value problem exists in at least a
neighbourhood of $\mathcal{S}_0$ if the spacetime and the initial data for $\xi_{AA'}$ are suitably smooth. 
Next, we define the spinor
$\mathring{S}_{AA'BB'}\equiv\nabla_{AA'}\mathring{\xi}_{BB'}+\nabla_{BB'}\mathring{\xi}_{AA'}$.
By a procedure similar to the one followed in the calculation of the
orthogonal decomposition of equation (\ref{Killing_eqn}), we work out
the orthogonal decomposition of $\mathring{S}_{AA'BB'}$ with respect
to the foliation ${\mathcal S}_s$ to obtain
\begin{eqnarray}
&&\mathring{S}_{CC'DD'}\tau_F^{\phantom{F}C'}\tau_L^{\phantom{L}D'}=
\mathring{\Sigma}_{DLCF}+\frac{1}{4}\epsilon_{CF}\epsilon_{DL}D\mathring{\xi}
+\frac{1}{2}\epsilon_{CF}\mathring{G}_{DL}+\frac{1}{2}\epsilon_{DL}\mathring{G}_{CF}. 
\label{s_ring}
\end{eqnarray}
with
\begin{eqnarray*}
&&\mathring{G}_{DL}\equiv-K_{DLAB}\mathring{\xi}^{AB}+
D_{DL}\mathring{\xi}+D\mathring{\xi}_{DL},\\ 
&&\mathring{\Sigma}_{DLCF}\equiv
D_{DL}\mathring{\xi}_{CF}+D_{CF}\mathring{\xi}_{DL}+\mathring{\xi} K_{CFDL}
\end{eqnarray*}
Equations (\ref{kid_1}) and (\ref{propagation:rule_a})-(\ref{propagation:rule_b})
 entail
\begin{displaymath}
\mathring{\Sigma}_{ABCD}|_{\mathcal{S}_0}=0, \quad
\mathring{G}_{AB}|_{\mathcal{S}_0}=0,
\end{displaymath} 
and thus
\begin{displaymath}
\mathring{S}_{AA'BB'}|_{\mathcal{S}_0}= \left.\left( \nabla_{AA'} \mathring{\xi}_{BB'} + \nabla_{BB'} \mathring{\xi}_{AA'} \right)\right|_{\mathcal{S}_0}=0.
\end{displaymath}
Next, it is noted that
\begin{equation}
\nabla_{CC'}\mathring{S}_{AA'BB'}= \frac{1}{2} \tau_{CC'} D\mathring{S}_{AA'BB'} -\updn{\tau}{Q}{C'} D_{CQ}\mathring{S}_{AA'BB'}+F_{CC'AA'BB'},
\label{nabla_ss}
\end{equation}
where $F_{CC'AA'BB'}$ is linear in $\mathring{S}_{AA'BB'}$.
Now, from $\mathring{S}_{AA'BB'}|_{\mathcal{S}_0}=0$, it follows that $D_{CD}\mathring{S}_{AA'BB'}|_{\mathcal{S}_0}=0$, and further that
$F_{CC'AA'BB'}|_{\mathcal{S}_0}=0$. On the other hand, applying the operator $D$ to equation (\ref{s_ring}) yields 
\begin{eqnarray}
&&\tau_F^{\phantom{F}C'}\tau_L^{\phantom{L}D'}D\mathring{S}_{CC'DD'}-
D\mathring{\Sigma}_{DLCF}=\nonumber\\
&&\hspace{2cm}=\frac{1}{2}\epsilon_{DF}\epsilon_{DL}D^2\mathring{\xi}+\frac{\epsilon_{DL}}{2}D\mathring{G}_{CF}+\frac{\epsilon_{CF}}{2}D\mathring{G}_{DL}.
\label{dS}
\end{eqnarray}
Equation (\ref{kid_2})
implies the property
\begin{equation}
D\mathring{\Sigma}_{DLCF}|_{\mathcal{S}_0}=\left.D(D_{DL}\mathring{\xi}_{CF}+D_{CF}\mathring{\xi}_{DL}+\mathring{\xi} K_{CFDL})\right|_{\mathcal{S}_0}=0,
\label{d_sigma}
\end{equation}
as can be readily checked by a calculation similar to the one which
enabled us to obtain equation (\ref{prekid_2}). Property
$\DAl\mathring{\xi}_{AA'}=0$ yields ---see the proof of proposition
\ref{proposition:Killing:propagation}---
\begin{displaymath}
\nabla_{AA'}\mathring{S}_{BB'}^{\phantom{BB'}AA'}-\frac{1}{2}
\nabla_{BB'}\mathring{S}^{AA'}_{\phantom{AA'}AA'}=0. 
\end{displaymath}
Using here (\ref{nabla_ss}) we get
after some calculations
\begin{eqnarray*}
&&\frac{1}{4}DS_{AA'}^{\phantom{AA'}AA'}+
\frac{1}{2}\tau^{AB'}\tau_{P}^{\phantom{P}A'}DS_{BA'AB'}-
\frac{1}{2}D_{BP}S_{AA'}^{\phantom{AA'}AA'}\\
&&
\hspace{3.5cm}-\tau^{AB'}\tau_{P}^{\phantom{P}A'}D_{A}^{\phantom{A}C}S_{BA'CB'}+F_{BP}=0, 
\end{eqnarray*}
where $F_{BP}$ is linear in $\mathring{S}_{AA'BB'}$.
If we replace here $\mathring{S}_{AA'BB'}$ by the expression given by (\ref{s_ring}) and 
evaluate the result on $\mathcal{S}_0$ we deduce 
\begin{displaymath}
\frac{1}{2}D\mathring{G}_{AB}|_{\mathcal{S}_0}+
\frac{1}{4}\epsilon_{AB}D^2\mathring{\xi}|_{\mathcal{S}_0}=0, 
\end{displaymath}
from which we conclude that $D\mathring{G}_{AB}|_{\mathcal{S}_0}=0$ and 
$D^2\mathring{\xi}|_{\mathcal{S}_0}=0$. Combining this with (\ref{dS}) and 
(\ref{d_sigma}) we obtain
that $D\mathring{S}_{AA'BB'}|_{\mathcal{S}_0}=0$ and thus due to the
above considerations
\begin{equation}
\nabla_{CC'}\mathring{S}_{AA'BB'}|_{\mathcal{S}_0}=\left.\nabla_{CC'}(\nabla_{AA'} \mathring{\xi}_{BB'} +\nabla_{BB'} \mathring{\xi}_{AA'})\right|_{{\mathcal S}_0}=0. 
\end{equation}
Finally, proposition \ref{proposition:Killing:propagation} now implies 
that $\mathring{S}_{AA'BB'}$ vanishes in a neighbourhood of
$\mathcal{S}_0$ so that $\mathring{\eta}_{AA'}$ is a Killing vector 
in that neighbourhood.\N

\begin{remark}
  \em From the previous discussion we deduce that equations
  (\ref{kid_1})-(\ref{kid_2}) are necessary and sufficient conditions
  for the existence of a Killing vector in the data development. These
  conditions are the (spinorial version of the) \emph{Killing initial
    data equations (KID equations)} ---see e. g.
  \cite{BeiChr97b,Col77} for a derivation of the tensorial version.
  These can be regarded as an overdetermined elliptic system for the
  \emph{longitudinal} ($\widetilde{\xi}$) and \emph{transverse} 
($\widetilde{\xi}_{AB}$) parts
  of the Killing vector $\xi_{AA'}$ at the initial hypersurface.
\end{remark}

\section{Twistor initial data}
\label{section:twistor_data}
Consider now a spinor of valence-1 satisfying the
twistor equation (\ref{twistor_eqn}), that is
\begin{displaymath}
\nabla_{A'(A}\kappa_{B)}=0.
\end{displaymath}
The spinor $\kappa_A$ is called a \emph{valence-1 Killing spinor}.
As mentioned in the introduction, a necessary condition for the
solvability of equation (\ref{twistor_eqn}) is that the spacetime be
of Petrov type N ---this can be seen by differentiating
(\ref{twistor_eqn}) and using the identities (\ref{box-relations}) to
render condition (\ref{typeN}).

If one is able to find conditions on a vacuum initial data set
ensuring the existence of a valence-1 Killing spinor in the data
development then one may conclude that the development must be of
Petrov type N.

\begin{proposition} \label{proposition:twistorpropagation}
Let $\kappa_A$ be a spinorial field on $\mathcal{M}$ such that
\begin{equation}
\nabla_{A'(A}\kappa_{B)}|_{\mathcal{S}_0}=0, \quad 
\left(\nabla_{EE'} \nabla_{A'(A}\kappa_{B)}\right)|_{\mathcal{S}_0}=0, \label{twistor:data1}
\end{equation}
where $\mathcal{S}_0\subset\mathcal{M}$ is a spacelike Cauchy hypersurface.
Assume further that 
\begin{displaymath}
\DAl \kappa_A=0
\end{displaymath}
 in an open set $\mathcal{W}$ which contains $\mathcal{S}_0$. 
Then, there exists an open set ${\mathcal U}\subset{\mathcal W}$ containing $\mathcal{S}_0$ such that the condition
\begin{displaymath}
\nabla_{A'(A} \kappa_{B)}=0,
\end{displaymath}
holds on $\mathcal U$.
\end{proposition}

\noindent
\textbf{Proof.}  Define
\begin{equation}
H_{A'AB} \equiv 2\nabla_{A'(A}\kappa_{B)}. \label{twistor:def}
\end{equation}
A straight-forward calculation using the commutators of the covariant derivatives renders
\begin{subequations}
\begin{eqnarray}
&&\DAl H_{A'AB} =2\nabla_{A'(A} \DAl \kappa_{B)} + 2 \dnup{\Psi}{AB}{PQ} H_{A'PQ}, 
\label{twistor:propagation1}\\
&&\DAl \kappa_A = \frac{2}{3} \nabla^{PP'} H_{P'PA}, \label{twistor:propagation2}
\end{eqnarray}
\end{subequations}
which in $\mathcal W$ becomes
\begin{equation}
\DAl H_{A'AB} = 2 \dnup{\Psi}{AB}{PQ} H_{A'PQ},\quad \nabla^{PP'} H_{P'PA}=0.
 \label{twistor:homogeneous}
\end{equation}
Now, (\ref{twistor:data1}) implies that  
\begin{equation}
H_{A'AB}|_{\mathcal{S}_0}=0, \quad \nabla_{EE'}H_{A'AB}|_{\mathcal{S}_0}=0. \label{twistor:homogeneous:data}
\end{equation}
Again, the initial value problem
(\ref{twistor:homogeneous})-(\ref{twistor:homogeneous:data}) has the
unique solution $H_{A'AB}=0$ in at least a neighbourhood 
$\mathcal{U}\subset\mathcal{W}$ containing $\mathcal{S}_0$. Hence $\kappa_A$ is a solution of the twistor equation on $\mathcal{U}$.\N
 
\begin{remark}
\em  As in the case of the proof of proposition
  \ref{proposition:Killing:propagation} the local character of the
  last proposition can be improved if the spacetime and
  $\kappa_A|_{\mathcal{S}_0}$ are suitably smooth.
\end{remark}

Next, we deduce intrinsic conditions on a vacuum initial data set
for its development to have a
valence-$1$ Killing spinor, and accordingly to be of Petrov type N.
The procedure is rather similar to that followed in section \ref{Kvector} for 
the case of Killing vectors. 
Suppose that $\kappa_{A}$ is a valence-1 Killing spinor. Then 
the orthogonal decomposition of the twistor condition with respect to 
the spinor $\tau^{AA'}$ gives
\begin{subequations}
\begin{eqnarray}
&& \nabla \kappa_B +\frac{2}{3} \dnup{\nabla}{B}{Q} \kappa_Q=0, \label{twistor:a}\\
&& \epsilon_{AC}\nabla \kappa_B + \epsilon_{BC} \nabla \kappa_A + 2 \nabla_{AC} \kappa_B + 2\nabla_{BC}\kappa_A=0. \label{twistor:b}
\end{eqnarray}
\end{subequations}
The symmetrisation of the last equation yields
\begin{equation}
\nabla_{(AB} \kappa_{C)}=0. \label{tid:equation1}
\end{equation}
It is also noted that the contraction of any two indices in
(\ref{twistor:b}) renders equation (\ref{twistor:a}). Hence the whole
content of the twistor equation is expressed in equations
(\ref{twistor:a}) and (\ref{tid:equation1}). Next, we use the relation between 
the operators $\nabla_{AB}$ and $D_{AB}$ to rewrite equation (\ref{tid:equation1}) in the form 
\begin{equation}
D_{(AB} \kappa_{C)}-\frac{1}{2} K_{Q(ABC)}\kappa^Q =0,
\label{twid:1}
\end{equation}
which is intrinsic to each of the integral surfaces $\mathcal{S}_s$ of
$\tau_{AA'}$.  This equation is known in older accounts as \emph{the
  spatial twistor equation} \cite{Tod84,Jef84b}.  From equation (\ref{twid:1})
we obtain the condition
\begin{displaymath}
D D_{(AB} \kappa_{C)}-\frac{1}{2}(\kappa^Q D K_{Q(ABC)}+ K_{Q(ABC)}D\kappa^Q) =0,
\end{displaymath}
which, by using the commutator relation (\ref{space_comm1}) and equation (\ref{evolutioneqn:K_ABCD}), can be transformed into
\begin{eqnarray} 
&&\hspace{-1cm} 
2D_{(AB}D_{F)C}\kappa^C+\frac{1}{2}\kappa_{(F}D_{AB)}K^{CD}_{\phantom{CD}CD}
-\frac{1}{2}K^{CD}_{\phantom{CD}CD}D_{(BF}\kappa_{A)}
-3\Omega_{CD(BF}D^{CD}\kappa_{A)}\nonumber\\
&&\hspace{1cm}+\frac{3}{2}\Omega_{(AF}^{\phantom{AF}DH}\Omega_{B)CDH}\kappa^C-
3(\mbox{i}B_{ABFC}+E_{ABFC})\kappa^C\nonumber\\
&&\hspace{1cm}+\frac{3}{4}K^{DH}_{\phantom{DH}DH}\Omega_{ABFC}\kappa^C-
\Omega_{ABFD}D_{C}^{\phantom{C}D}\kappa^C=0. 
\label{twid:2}
\end{eqnarray}
where we have defined $\Omega_{ABCD}\equiv K_{(ABCD)}$, the trace-free
part of the second fundamental form.

Equations (\ref{tid:equation1}) and (\ref{twid:2}) are intrinsic to
each of the leaves $\mathcal{S}_s$ and therefore they are necessary
conditions for the existence of a twistor in the development of a
vacuum initial data set. The converse of this statement also holds and
its proof is analogous to that of proposition
\ref{proposition:Killing:data}.

\begin{proposition}[\bf{Twistor initial data}]
  \label{proposition:twistor:data} 
Let $(\mathcal{S},h_{ij},K_{ij})$
  be an initial data set for the vacuum Einstein field equations and
  let $({\mathcal M},g_{\mu\nu})$ be its data development. Assume
  further that on $\mathcal S$ there exists a spinor
  $\widetilde{\kappa}_A$ satisfying the conditions (twistor initial
  data equations)
\begin{subequations}
\begin{eqnarray}
&& D_{(AB}\widetilde{\kappa}_{C)}+\frac{1}{2}
\widetilde{K}^{Q}_{\phantom{Q}(ABC)}\widetilde{\kappa}_Q =0,
\label{twistorid_1}\\
&&2 D_{(AB}D_{F)C}\widetilde{\kappa}^C+\frac{1}{2}\widetilde{\kappa}_{(F}D_{AB)}
\widetilde{K}^{CD}_{\phantom{CD}CD}
-\frac{1}{2}\widetilde{K}^{CD}_{\phantom{CD}CD}D_{(BF}\widetilde{\kappa}_{A)}
\nonumber \\
&&\hspace{1.5cm}-3\widetilde{\Omega}_{CD(BF}D^{CD}\widetilde{\kappa}_{A)}+
\frac{3}{2}\widetilde{\Omega}_{(AF}^{\phantom{AF}DH}\widetilde{\Omega}_{B)CDH}\widetilde{\kappa}^C-
3(\mbox{\em i}\widetilde{B}_{ABFC}+\widetilde{E}_{ABFC})\widetilde{\kappa}^C\nonumber\\
&&\hspace{1.5cm}+\frac{3}{4}\widetilde{K}^{DH}_{\phantom{DH}DH}
\widetilde{\Omega}_{ABFC}\widetilde{\kappa}^C+
\widetilde{\Omega}_{ABF}^{\phantom{ABF}D}D_{CD}\widetilde{\kappa}^C=0
\label{twistorid_2}
\end{eqnarray}
\end{subequations}
Then there exists a spinorial field $\mathring{\kappa}_{A}$ satisfying
the twistor equation (\ref{twistor_eqn}) on an open subset of
$\mathcal{S}$.
\end{proposition}
\noindent
\textbf{Proof.} Consider a foliation of the development $\mathcal M$
with leaves $\mathcal{S}_s$. Identify $\mathcal{S}_0$ with $\mathcal
S$. In a neighbourhood of $\mathcal{S}_0$ this foliation is chosen in
such a way that $K_{AB}=0$. We introduce a spacetime spinor
$\mathring{\kappa}_A$ ---the Killing spinor candidate--- satisfying
\begin{displaymath}
\DAl \mathring{\kappa}_A=0,
\end{displaymath}
subject to the initial conditions
\begin{subequations}
\begin{eqnarray}
&&\mathring{\kappa}_A|_{\mathcal{S}_0}=\widetilde{\kappa}_A, \label{twistor_propagation_a}\\
&& D\mathring{\kappa}_A|_{\mathcal{S}_0}=\frac{1}{6}K^{CF}_{\phantom{CF}CF}\widetilde{\kappa}_A-
\frac{2}{3}D_{A}^{\phantom{A}C}\widetilde{\kappa}_C.\label{twistor_propagation_b}
\end{eqnarray}
\end{subequations}
Again, theorem \ref{hyperbolic_propagation} guarantees that 
the above initial value problem has a solution in at least a
neighbourhood of $\mathcal{S}_0$ if the spacetime and initial data 
are suitably smooth. 
Define $\mathring{H}_{A'AB}\equiv
2\nabla_{A'(A}\mathring{\kappa}_{B)}$.  A computation similar to that
carried out to obtain equations (\ref{twistor:a})-(\ref{twistor:b}) gives
\begin{equation}
\mathring{H}_{A'AB}\tau^{A'}_{\phantom{A'}D}=
2\mathring{\Sigma}_{ABD}+
\epsilon_{BD}\mathring{G}_A+\frac{1}{2}\epsilon_{AB}\mathring{G}_D,
\label{htwistor_decomposition}
\end{equation}
with 
\begin{displaymath}
\mathring{\Sigma}_{ABD}\equiv
\left(D_{(AB}\mathring{\kappa}_{D)}-\frac{1}{2}K_{(ABD)C}\mathring{\kappa}^C\right),\quad
\mathring{G}_{A}\equiv
\left(D\mathring{\kappa}_A-\frac{1}{6}K^{CF}_{\phantom{CF}CF}\mathring{\kappa}_A-\frac{2}{3}D_{AC}\mathring{\kappa}^C\right).
\end{displaymath}
From (\ref{twistor_propagation_a})-(\ref{twistor_propagation_b}) we
deduce $\mathring{G}_A|_{\mathcal{S}_0}=0$ and (\ref{twistorid_1})
yields $\mathring{\Sigma}_{ABC}|_{\mathcal{S}_0}=0$. Therefore
\begin{displaymath}
\mathring{H}_{A'AB}|_{\mathcal{S}_0}=2\nabla_{A'(A} \mathring{\kappa}_{B)}|_{\mathcal{S}_0}=0.
\end{displaymath}
Next, we consider the relation 
\begin{equation}
\tau_{P}^{\phantom{P}C'}\nabla_{CC'}\mathring{H}_{A'AB}=D_{CP}\mathring{H}_{A'AB}+
\frac{1}{2}\epsilon_{CP}D\mathring{H}_{A'AB}+F_{ABCPA'},
\label{cd_hAAB}
\end{equation}
where $F_{ABCPA'}$ is linear in $\mathring{H}_{A'AB}$. Since we have
$\mathring{H}_{A'AB}|_{\mathcal{S}_0}=0$ we deduce
$D_{CP}\mathring{H}_{A'AB}|_{\mathcal{S}_0}=0$. On the other hand
applying the operator $D$ to equation (\ref{htwistor_decomposition})
we obtain
\begin{equation}
\tau_D^{\phantom{D}A'}D\mathring{H}_{A'AB}-
2D\mathring{\Sigma}_{ABD}=
\epsilon_{BD}D\mathring{G}_A+\frac{1}{2}\epsilon_{AB}D\mathring{G}_D,
\label{dh}
\end{equation}
 From condition
(\ref{twistorid_2}) we deduce
\begin{displaymath}
D\mathring{\Sigma}_{ABD}|_{\mathcal{S}_0}=\left.D\left(D_{(AB}\mathring{\kappa}_{D)}
-\frac{1}{2}K_{(ABD)C}\mathring{\kappa}^C\right)\right|_{\mathcal{S}_0}=0,
\end{displaymath}
with an argument similar to that of the calculation which led to
(\ref{twid:2}). Also the condition $\DAl\mathring{\kappa}_{AB}=0$
implies
\begin{displaymath}
 \nabla^{PP'}\mathring{H}_{P'PA}=0,
\end{displaymath}
as is clear from the proof of proposition \ref{proposition:twistorpropagation}.
Using in this expression (\ref{cd_hAAB}) and (\ref{dh}) we get
\begin{eqnarray*}
&&\frac{3}{4}D\mathring{G}_A+2D_{PB}\mathring{\Sigma}_{A}^{\phantom{A}PB}+
\frac{1}{2}D_{AP}\mathring{G}^P-\\
&&-\frac{1}{4}\mathring{G}^PK_{A\phantom{B}PB}^{\phantom{A}B}+\mathring{G}_A
K^{PB}_{\phantom{PB}PB}-K_{APBC}\mathring{\Sigma}^{PBC}=0, 
\end{eqnarray*}
from which we conclude $D\mathring{G}_{AB}|_{\mathcal{S}_0}=0$
and hence, via  (\ref{dh}),
$D\mathring{H}_{A'AB}|_{\mathcal{S}_0}=0$. Thus
(\ref{cd_hAAB}) yields
\begin{equation}
\nabla_{CC'}\mathring{H}_{A'AB}|_{\mathcal{S}_0}=
\nabla_{CC'}\nabla_{A'(A}\mathring{\kappa}_{B)}|_{\mathcal{S}_0}=0.
\end{equation}
Proposition \ref{proposition:twistorpropagation} implies that
$\mathring{H}_{A'AB}=0$ in a neighbourhood of $\mathcal{S}_0$ and
hence $\mathring{\kappa}_A$ is a solution of the twistor equation
(\ref{twistor_eqn}) in such a neighbourhood.\N

An application of the last result is the following 
\begin{theorem}[Type N initial data] \label{typeN:data}
Let $(\mathcal{S},h_{ij},K_{ij})$ be a vacuum initial data set and suppose that there exists a spinor $\widetilde{\kappa}_A$ on $\mathcal{S}$ fulfilling (\ref{twistorid_1})-(\ref{twistorid_2}).
Then the development $(\mathcal{M},g_{\mu\nu})$ contains an open set $\mathcal{W}$ such that $(\mathcal{W},g_{\mu\nu})$ is of Petrov type N.
\end{theorem}
\noindent
\textbf{Proof.} This is a direct consequence of equation (\ref{typeN}) and of proposition \ref{proposition:twistor:data}.\N 

\section{Valence-$2$ Killing spinor initial data}
\label{section:killing_spinor_data}
Next, we explain how to construct vacuum initial data such that their 
development contains a valence-$2$ Killing spinor. 
As we did in the case of Killing vectors and twistors, we start by finding
a hyperbolic system which will be used as the basis to study the 
\emph{propagation} of the differential condition which guarantees the existence of valence-$2$ Killing
spinors. In order to obtain such a system of
propagation equations, one has to consider simultaneously the propagation
of the Killing vector associated to the Killing spinor.

\begin{proposition} \label{proposition:Kspinorpropagation} Let
  $\kappa_{AB}$ be a spinor defined on $\mathcal M$ and such that on a
  spacelike Cauchy hypersurface $\mathcal{S}_0$ one has
\begin{subequations}
\begin{eqnarray}
&& \nabla_{A'(A}\kappa_{BC)}|_{\mathcal{S}_0}=0, \label{Kspinor:data1}\\
&& \nabla_{EE'} \nabla_{A'(A}\kappa_{BC)}|_{\mathcal{S}_0}=0, \label{Kspinor:data2}\\
&&\left.\left(\nabla_{AA'} \updn{\nabla}{P}{B'}\kappa_{BP} + \nabla_{BB'} \updn{\nabla}{P}{A'} \kappa_{AP}\right)\right|_{\mathcal{S}_0}=0, \label{Kspinor:data3}\\
&& \left.\nabla_{EE'} \left(\nabla_{AA'} \updn{\nabla}{P}{B'}\kappa_{BP} + \nabla_{BB'} \updn{\nabla}{P}{A'}\kappa_{AP}\right)\right|_{\mathcal{S}_0}=0 \label{Kspinor:data4}.
\end{eqnarray}
\end{subequations}
Assume further that the condition 
$$
\DAl\kappa_{AB}+\Psi_{ABPQ}\kappa^{PQ}=0
$$ 
holds on an open set $\mathcal{W}$
containing $\mathcal{S}_0$.
Then there exists an open set $\mathcal{U}\subset\mathcal{M}$ containing 
$\mathcal{S}_0$ such that
\begin{displaymath}
\nabla_{A'(A} \kappa_{BC)}=0, 
\end{displaymath}
on $\mathcal U$.
\end{proposition}

\textbf{Proof.} 
Define
\begin{subequations}
\begin{eqnarray}
&&H_{A'ABC} \equiv 3\nabla_{A'(A} \kappa_{BC)}, \label{start-entry_a}\\
&&\xi_{AA'}\equiv\nabla^{D}_{\phantom{D}A'}\kappa_{DA},\label{start-entry_b} \\
&&S_{AA'BB'} \equiv \nabla_{AA'} \xi_{BB'} + \nabla_{BB'} \xi_{AA'}.
\label{start-entry_c}
\end{eqnarray}
\end{subequations}
As in previous sections, the general strategy will be to construct a
hyperbolic system with $H_{A'ABC}$ as one of its unknowns. First of
all, we need to find a relation between $S_{AA'BB'}$ and $H_{A'ABC}$.
To that end, we replace in (\ref{start-entry_c}) the spinor $\xi_{AA'}$ by its expression in
terms of $\kappa_{AB}$. This gives
\begin{equation}
S_{CC'DD'}=-\nabla_{DD'}\nabla^{A}_{\phantom{A}C'}\kappa_{CA}-
\nabla_{CC'}\nabla^{A}_{\phantom{A}D'}\kappa_{DA}. 
\label{s1}
\end{equation}
Now, in this last expression we use the identity
\begin{eqnarray}
&&2\nabla_{DD'}\nabla^{A}_{\phantom{A}C'}\kappa_{CA}=
-\epsilon_{CD}\overline{\epsilon}_{C'D'}\nabla^{B}_{\phantom{B}A'}\nabla^{AA'}\kappa_{AB}
+2\nabla_{C(D'}\nabla^{A}_{\phantom{A}C')}\kappa_{DA}\nonumber\\
&&\hspace{4cm}+2\epsilon_{CD}\nabla^{A}_{\phantom{A}(C'}\nabla^{B}_{\phantom{B}D')}\kappa_{AB}
+\overline{\epsilon}_{C'D'}\nabla_{C}^{\phantom{C}A'}\nabla^A_{\phantom{A}A'}\kappa_{DA}.
\label{identiy_2}
\end{eqnarray}
After some lengthy algebra involving the commutation of the covariant
derivatives and the grouping of some terms by means of
$H_{A'ABC}=3\nabla_{A'(A} \kappa_{BC)}$, we arrive at ---see appendix
B for further details about this calculation---
\begin{equation}
S_{CC'DD'}=-\frac{1}{2}\nabla^A_{\phantom{A}C'}H_{D'CDA},
\label{covdiv_H}
\end{equation}
which is kept for later use. 
A straight-forward
calculation using the decomposition of a spinor in terms of its
totally symmetric part and symmetrised contractions yields
\begin{displaymath}
\nabla_{EE'} H_{A'ABC}= \nabla_{E'(E} H_{ABC)A'} +\frac{1}{2}(\epsilon_{EA} S_{BE'CA'} +\epsilon_{EB}S_{AE'CA'}+\epsilon_{EC}S_{AE'BA'}),
\end{displaymath}
where equation (\ref{covdiv_H}) has been used. Now, using that
\begin{displaymath}
\nabla_{E'(E} H_{ABC)A'}=\frac{1}{4} \left(\nabla_{E'E} H_{ABCA'} + \nabla_{E'A} H_{EBCA'} + \nabla_{E'B} H_{EACA'} + \nabla_{E'C} H_{EABA'}  \right), 
\end{displaymath}
we obtain
\begin{eqnarray}
&&\hspace{-1cm}\DAl H_{A'ABC}= \frac{1}{4} \DAl H_{A'ABC} \nonumber\\
&&\hspace{2cm}+\frac{1}{4}\left(\nabla^{EE'}\nabla_{E'A} H_{A'EBC} + \nabla^{EE'}\nabla_{E'B} H_{A'EAC} +\frac{1}{4}\nabla^{EE'}\nabla_{E'B} H_{A'EAC} \right) \nonumber\\
&&\hspace{2cm}+\frac{1}{2}\left(\dnup{\nabla}{A}{E'} S_{BE'CA'} + \dnup{\nabla}{B}{E'} S_{AE'CA'} +\dnup{\nabla}{C}{E'}S_{AE'BA'}\right).
\label{pre_hpropagation}
\end{eqnarray}
Now, if we make use of the identity
\begin{equation}
\nabla_{AC'}\nabla_{B}^{\phantom{B}C'}=
 \DAl_{AB}+\frac{1}{2}\epsilon_{AB}\DAl,
\label{identity:nablas}
\end{equation}
and of expression (\ref{box-relations}), equation (\ref{pre_hpropagation}) reduces to
\begin{equation}
\DAl H_{A'ABC}= 4 (\dnup{\Psi}{(AB}{PQ} H_{C)PQA'} +\dnup{\nabla}{(A}{Q'}
S_{BC)Q'A'}). \label{Kspinor:propagation}
\end{equation}
The latter hyperbolic equation for $H_{A'ABC}$ has to be complemented
with another hyperbolic equation for $S_{AA'BB'}$ which we compute next. 
We calculate $\nabla^{CA'}H_{A'ABC}$ from equation (\ref{start-entry_a}) to obtain 
\begin{displaymath}
\nabla^{CA'}H_{A'ABC}=\DAl\kappa_{AB}-2
\nabla^{C}_{\phantom{C}A'}\nabla_{(B}^{\phantom{B}A'}\kappa_{A)C}.
\end{displaymath}
We work out the last term of the right hand side with equations
(\ref{identity:nablas}) and (\ref{box-relations}). 
The final result is
\begin{equation}
\nabla^{CA'}H_{A'ABC}=\DAl \kappa_{AB}+\Psi_{ABPQ} \kappa^{PQ}=0. \label{Box_kappa_AB}
\end{equation}
where, by hypothesis, the last equality only holds in the open set $\mathcal{W}$. 
Also a direct calculation shows that in $\mathcal{W}$
\begin{equation}
\DAl \xi_{AA'} = \DAl \updn{\nabla}{Q}{A'}\kappa_{AQ}= -\dnup{\Psi}{A}{PQR} H_{A'PQR}, \label{Box_xi}
\end{equation}
where again $\DAl \kappa_{AB}+\Psi_{ABPQ} \kappa^{PQ}=0$ was used. 
Hence, substitution of (\ref{Box_xi}) into equation
(\ref{Killing:propagation1}) yields
\begin{eqnarray}
&& \DAl S_{AA'BB'} = -\nabla_{AA'} (\dnup{\Psi}{B}{PQR}H_{B'PQR}) -\nabla_{BB'}(\dnup{\Psi}{A}{PQR}H_{A'PQR}) \nonumber \\
&&\hspace{3.5cm}+ 2 \dnup{\Psi}{AB}{PQ} S_{PA'QB'} + 2\dnup{\overline{\Psi}}{A'B'}{P'Q'} S_{AP'BQ'}, \label{Kvector:propagation}
\end{eqnarray}
which is the required hyperbolic equation for $S_{AA'BB'}$.  Now, we
note that the system of hyperbolic partial differential equations
formed by equations (\ref{Kspinor:propagation}) and
(\ref{Kvector:propagation}) falls within theorem
\ref{hyperbolic_propagation} if we take $H_{A'ABC}$ and $S_{AA'BB'}$
as the unknowns. Conditions
(\ref{Kspinor:data1})-(\ref{Kspinor:data4}) imply that the initial
data of such a hyperbolic system are
\begin{eqnarray*}
&& H_{A'ABC}|_{\mathcal{S}_0}=0,\quad \nabla_{DD'}H_{A'ABC}|_{\mathcal{S}_0}=0,\\
&& S_{AA'BB'}|_{\mathcal{S}_0}=0,\quad \nabla_{DD'}S_{AA'BB'}|_{\mathcal{S}_0}=0,
\end{eqnarray*}
and therefore we deduce that $H_{A'ABC}=0$ and $S_{AA'BB'}=0$ in a
neighbourhood of $\mathcal{S}_0$. Note that equations (\ref{covdiv_H}),
(\ref{Box_kappa_AB})
which are constraints of the dependent variables in the hyperbolic system are
now trivially fulfilled. \N

\begin{remark}
\em  Again, if the spacetime and the restriction of $\kappa_{AB}$ to
  $\mathcal{S}_0$ are suitably smooth, then it is possible to extend
  the existence of the Killing spinor to the whole of $\mathcal{M}$.
\end{remark}

\medskip
From the proof of this proposition we note the following 
\begin{corollary} \label{corollary:propagationreality}
If $\mathring{\xi}_{AA'}=\updn{\nabla}{Q}{A'} \mathring{\kappa}_{QA}$ is such
that $\overline{\mathring{\xi}}_{AA'}=\mathring{\xi}_{AA'}$, $\nabla_{CC'}\overline{\mathring{\xi}}_{AA'}=\nabla_{CC'}\mathring{\xi}_{AA'}$ on $\mathcal{S}_0$
then $\overline{\xi}_{AA'}=\xi_{AA'}$ on $\mathcal{M}$.  
\end{corollary}

\textbf{Proof.} This follows directly from the wave equation
(\ref{Box_xi}). \hfill $\Box$

\medskip Hence, if the Killing vector associated to the Killing vector
is real on the initial hypersurface, then it is also real at later
times. This corollary is useful to characterise a class of initial
data sets for Petrov type D spacetimes which includes initial data
sets for the Kerr spacetime ---see section \ref{section:typeD}.

Next, we proceed to obtain necessary conditions for the development of a
vacuum initial data set to admit valence-$2$ Killing spinors. This is
accomplished in a similar fashion as in the previous sections.  The
orthogonal decomposition of the Killing spinor equation
(\ref{Kspinor_eqn}) renders the expressions
\begin{subequations}
\begin{eqnarray}
&& \nabla \kappa_{BC} + \nabla_{(B}^{\phantom{(B}A} \kappa_{C)A}=0, \label{ks1}\\
&& \epsilon_{AD}\nabla \kappa_{BC} +\epsilon_{CD}\nabla \kappa_{AB}+ \epsilon_{BD}\nabla \kappa_{AC} \nonumber\\
&&\hspace{2cm}+2( \nabla_{AD} \kappa_{BC} + \nabla_{CD} \kappa_{AB} + \nabla_{BD} \kappa_{AC})=0. \label{ks2}
\end{eqnarray}
\end{subequations}
The $\nabla$-derivative in (\ref{ks1}) can be transformed
into a $D$-derivative to yield 
\begin{equation}
 D\kappa_{AC}=-D_{(C}^{\phantom{C}B}\kappa_{A)B}
-\frac{1}{4}K^{BF}_{\phantom{BF}BF}\kappa_{AC}
-\frac{1}{2}K_{ABCF}\kappa^{BF}-K_{(C}^{\phantom{C}B}\kappa_{A)B}.
\label{ks5}
\end{equation}
Hence equation (\ref{ks2}) is equivalent to
\begin{displaymath}
\nabla_{(AB} \kappa_{CD)}=0
\end{displaymath}
so if we transform in this expression 
the covariant derivative $\nabla_{AB}$ into $D_{AB}$ we obtain. 
\begin{equation}
D_{(AB} \kappa_{CD)}-K_{E(ABC} \dnup{\kappa}{D)}{E}=0. \label{ks4}
\end{equation}
Equations (\ref{ks5})-(\ref{ks4}) are completely equivalent to
(\ref{ks1})-(\ref{ks2}). Equation (\ref{ks4}) is intrinsic to the
leaves $\mathcal{S}_s$ and hence it is a necessary condition for the
existence of a valence-2 Killing spinor in the data development.
Another necessary condition is obtained from
\begin{displaymath}
D(D_{(AB} \kappa_{CD)}-K_{E(ABC} \dnup{\kappa}{D)}{E})=0. 
\end{displaymath}
As in previous sections we transform this equation by means of the
commutator (\ref{space_comm1}), equation (\ref{evolutioneqn:K_ABCD})
and equation (\ref{ks5}). We choose a foliation with vanishing
acceleration in at least a neighbourhood of $\mathcal{S}_0$
in order to perform these calculations. The resulting expression is
\begin{eqnarray*}
&&\hspace{-1cm}D_{(AC}D_{B}^{\phantom{B}D}\kappa_{F)D}+\frac{1}{2}D_{(AB}(\kappa^{DH}\Omega_{CF)DH})+
\Omega_{DH(BF}D^{DH}\kappa_{AC)}\nonumber\\
&&-\frac{1}{2}\Omega_{H(ABC}D_{F)}^{\phantom{F}D}\kappa_{D}^{\phantom{D}H}
-\frac{1}{2}\Omega_{H(ACF}D^{DH}\kappa_{B)D}+
2(\mbox{i}B_{D(BCF}+E_{D(BCF})\kappa_{A)}^{\phantom{A}D}\nonumber\\
&&-\left(\frac{1}{3}K^{HL}_{\phantom{HL}HL}\Omega_{D(ABC}+
\Omega_{DHL(A}\Omega_{BC}^{\phantom{BC}HL}\right)\kappa_{F)}^{\phantom{F}D}+
\frac{1}{2}\kappa^{DH}\Omega_{DHL(A}\Omega_{BCF)}^{\phantom{BCF}L}\nonumber\\
&&-\frac{1}{3}\kappa_{(AB}D_{CF)}K^{DH}_{\phantom{DH}DH}=0. \label{ks7}
\end{eqnarray*}
Another set of necessary conditions arises from the orthogonal
decomposition of $\xi_{FA'}=\nabla^D_{\phantom{D}A'}\kappa_{DF}$. In
the spirit of the space-spinor formalism we write again
\begin{displaymath}
 \xi_{AA'}=\frac{1}{2}\xi\tau_{AA'}-\updn{\tau}{Q}{A'}\xi_{AQ}.
\end{displaymath}
A direct calculation shows that
\begin{subequations}
\begin{eqnarray}
&& \xi =\xi_{AA'} \tau^{AA'}=\nabla^{PQ} \kappa_{PQ}=D^{PQ}\kappa_{PQ},\label{xi_1}\\
&& \xi_{AB}=\dnup{\tau}{(A}{C'} \xi_{B)C'}= 
\updn{\nabla}{Q}{(A}\kappa_{B)Q} -\frac{1}{2} \nabla 
\kappa_{AB} \nonumber\\
&&\phantom{ \xi_{AB}}=-\frac{1}{2}K^{PQ}_{\phantom{PQ}PQ}\kappa_{AB}+
\frac{3}{4}\kappa^{PQ}\Omega_{ABPQ}+\frac{3}{2}D_{(A}^{\phantom{A}P}\kappa_{B)P}
\label{xi_2}
\end{eqnarray}
\end{subequations}
where in the last equation the propagation equation (\ref{ks5}) has
been used to simplify. 

As it is to be expected, one has the following result 

\begin{proposition}[Valence-2 Killing spinor initial data]
  \label{proposition:Kspinor:data}
  Let $(\mathcal{S},h_{ij},K_{ij})$ be an initial data set for the
  vacuum Einstein field equations such that there exists a
  space spinor $\widetilde{\kappa}_{AB}$ on $\mathcal{S}$ satisfying
  the equations
\begin{eqnarray}
&& D_{(AB}\widetilde{\kappa}_{CD)}-\widetilde{K}_{E(ABC} \dnup{\widetilde{\kappa}}{D)}{E}=0,
\label{killingspinor_a}\\ 
&&D_{(AC}D_{B}^{\phantom{B}D}\widetilde{\kappa}_{F)D}+
\frac{1}{2}D_{(AB}(\widetilde{\kappa}^{DH}\widetilde{\Omega}_{CF)DH})+
\widetilde{\Omega}_{DH(BF}D^{DH}\widetilde{\kappa}_{AC)}\nonumber\\
&&\hspace{1cm}-\frac{1}{2}\widetilde{\Omega}_{H(ABC}D_{F)}^{\phantom{F}D}
\widetilde{\kappa}_{D}^{\phantom{D}H}
-\frac{1}{2}\widetilde{\Omega}_{H(ACF}D^{DH}\widetilde{\kappa}_{B)D}+
2(\mbox{\em i}\widetilde{B}_{D(BCF}+\widetilde{E}_{D(BCF})
\widetilde{\kappa}_{A)}^{\phantom{A}D}\nonumber\\
&&\hspace{1cm}-\left(\frac{1}{3}\widetilde{K}^{HL}_{\phantom{HL}HL}
\widetilde{\Omega}_{D(ABC}+
\widetilde{\Omega}_{DHL(A}\widetilde{\Omega}_{BC}^{\phantom{BC}HL}\right)\widetilde{\kappa}_{F)}^{\phantom{F}D}+\frac{1}{2}\widetilde{\kappa}^{DH}
\widetilde{\Omega}_{DHL(A}\widetilde{\Omega}_{BCF)}^{\phantom{BCF}L}\nonumber\\
&&\hspace{1cm}-\frac{1}{3}\widetilde{\kappa}_{(AB}D_{CF)}\widetilde{K}^{DH}_{\phantom{DH}DH}=0.
\label{killingspinor_b}
\end{eqnarray}
In addition, assume that the space spinors 
$\widetilde{\xi}$, $\widetilde{\xi}_{BF}$ defined by
\begin{eqnarray}
&&\widetilde{\xi}\equiv D^{PQ}\widetilde{\kappa}_{PQ},\label{tildexi_1} \\
&&\widetilde{\xi}_{BF}\equiv-\frac{1}{2}\widetilde{K}^{DA}_{\phantom{DA}DA}
\widetilde{\kappa}_{BF}+
\frac{3}{4}\widetilde{\kappa}^{DA}\widetilde{\Omega}_{BFDA}+
\frac{3}{2}D_{(F}^{\phantom{F}D}\widetilde{\kappa}_{B)D},\label{tildexi_2}
\end{eqnarray}
are such that they fulfil the conditions (\ref{kid_1})-(\ref{kid_2})
of proposition \ref{proposition:Killing:data}.  Then, there exists a
spacetime spinor $\mathring{\kappa}_{AB}$ in a neighbourhood of the
data development $\mathcal M$ which is a valence-2 Killing spinor.
\label{killing_spinor_initial_data}
\end{proposition}

\textbf{Proof.} The proof of this result proceeds in a similar way as
the proofs of propositions \ref{proposition:Killing:data} and
\ref{proposition:twistor:data}.  We consider a foliation of the data
development $\mathcal M$ whose leaves are $\mathcal{S}_s$. Identify 
$\mathcal{S}_0$ with $\mathcal{S}$. In a neighbourhood of
$\mathcal{S}_0$ the foliation is constructed in such a way that
$K_{AB}=0$. We consider a Killing spinor candidate
$\mathring{\kappa}_{AB}$ satisfying
\begin{displaymath}
\DAl \mathring{\kappa}_{AB}= -\Psi_{ABPQ} \mathring{\kappa}^{PQ},
\end{displaymath}
with initial data on $\mathcal{S}_0$ given by
\begin{subequations}
\begin{eqnarray}
&&\mathring{\kappa}_{AB}|_{\mathcal{S}_0}=\widetilde{\kappa}_{AB}, \label{ind_kappa_a}\\
&&D\mathring{\kappa}_{AC}|_{\mathcal{S}_0}=-D_{(C}^{\phantom{C}B}\widetilde{\kappa}_{A)B}-
\frac{1}{4}\widetilde{K}^{BF}_{\phantom{BF}BF}\widetilde{\kappa}_{AC}
-\frac{1}{2}\widetilde{K}_{ABCF}\widetilde{\kappa}^{BF}, \label{ind_kappa_b}
\end{eqnarray}
\end{subequations}
Again, theorem \ref{hyperbolic_propagation} ensures that this initial value problem has a solution in at least a neighbourhood of
$\mathcal{S}_0$ if the spacetime and the initial data for the Killing
spinor are suitably smooth. Next we define
$\mathring{H}_{A'ABC}\equiv3\nabla_{A'(A}\mathring{\kappa}_{BC)}$ and
compute its orthogonal decomposition by a procedure similar to that
followed to obtain the relations (\ref{ks5})-(\ref{ks4}). This renders
\begin{equation}
 \mathring{H}_{A'ABC}\tau_F^{\phantom{F}A'}=3\mathring{\Sigma}_{ABCF}+
\frac{1}{8}(\epsilon_{CF}\mathring{G}_{AB}+
\epsilon_{AB}\mathring{G}_{CF})+\frac{1}{4}\epsilon_{BF}\mathring{G}_{AC},
\label{killingspinor_1st}
\end{equation}
where
\begin{eqnarray*}
&&\mathring{G}_{AB}\equiv-K^{DH}_{\phantom{DH}DH}\mathring{\kappa}_{AB}+
2K_{ADBH}\mathring{\kappa}^{DH}+
4D_{(B}^{\phantom{B}D}\mathring{\kappa}_{A)D}+4D\mathring{\kappa}_{AB},\\ 
&&\mathring{\Sigma}_{ABCF}\equiv(D_{(AB}\mathring{\kappa}_{CF)}-
K_{E(ABC}\dnup{\mathring{\kappa}}{F)}{E})
\end{eqnarray*}
Clearly, conditions (\ref{ind_kappa_b}) and (\ref{killingspinor_a}) entail
$\mathring{G}_{AB}|_{\mathcal{S}_0}=0$ and $\mathring{\Sigma}_{ABCD}|_{\mathcal{S}_0}=0$
respectively from which we get
\begin{displaymath}
\mathring{H}_{A'ABC}|_{\mathcal{S}_0}=
\nabla_{A'(A}\mathring{\kappa}_{BC)}|_{\mathcal{S}_0}=0.
\end{displaymath}
In addition, we have 
\begin{equation}
\nabla_{AA'}\mathring{H}_{B'BCD}=\frac{1}{2}\tau_{AA'}D\mathring{H}_{B'BCD}-
\tau^F_{\phantom{F}A'}D_{AF}\mathring{H}_{B'BCD}+F_{AA'B'BCD},
\label{cd_Hbbcd}
\end{equation}
with $F_{AA'B'BCD}$ linear in $\mathring{H}_{B'BCD}$. Therefore, from
the above $D_{AB}\mathring{H}_{A'CDF}|_{\mathcal{S}_0}=0$ and
$F_{AA'B'BCD}|_{\mathcal{S}_0}=0$.  On the other hand, if we apply the
operator $D$ to equation (\ref{killingspinor_1st}) we obtain 
\begin{equation}
\tau_F^{\phantom{F}A'}D\mathring{H}_{A'ABC}-
3D\mathring{\Sigma}_{ABCF}=
\frac{1}{4}\epsilon_{BF}D\mathring{G}_{AC}+
\frac{1}{8}(\epsilon_{CF}D\mathring{G}_{AB}+\epsilon_{AB}D\mathring{G}_{CF}),
\label{dt_HABC}
\end{equation}
Condition (\ref{killingspinor_b}) entails   
\begin{displaymath}
D\mathring{\Sigma}_{ABCD}|_{\mathcal{S}_0}=\left.D\left(D_{(AB}\mathring{\kappa}_{CF)}-K_{E(ABC}\dnup{\mathring{\kappa}}{F)}{E}\right)\right|_{\mathcal{S}_0}=0,
\end{displaymath}
as it is shown by a computation similar to the one which enabled us to obtain 
equation (\ref{ks7}). Also the condition $\DAl\mathring{\kappa}_{AB}+\Psi_{ABCD}\mathring{\kappa}^{CD}=0$ implies
 ---cfr. equation (\ref{Box_kappa_AB})---
$$
\nabla^{CA'}\mathring{H}_{A'ABC}=0.
$$
We work out the orthogonal splitting of this condition by using (\ref{cd_Hbbcd}) and (\ref{killingspinor_1st}) with the result
\begin{eqnarray*}
 \frac{1}{4}D\mathring{G}_{AB}-\frac{1}{4}D_B^{\phantom{B}C}\mathring{G}_{AC}+
\frac{1}{8}\epsilon_{AB}D^{CD}\mathring{G}_{CD}+3D^{CD}\mathring{\Sigma}_{ABCD}+\\
+\frac{1}{3}\mathring{G}_{AB}K^{CD}_{\phantom{CD}CD}-
\frac{1}{8}\mathring{G}^{CD}\mathring{\Omega}_{ABCD}-
3\mathring{\Sigma}_{(B}^{\phantom{B}CDF}\mathring{\Omega}_{A)CDF}=0,
\end{eqnarray*}
from which we deduce that $D\mathring{G}_{AB}|_{\mathcal{S}_0}=0$.
Thus (\ref{dt_HABC}) implies $D\mathring{H}_{A'ABC}|_{\mathcal{S}_0}=0$ and hence 
(\ref{cd_Hbbcd}) yields
\begin{displaymath}
\nabla_{AA'}\mathring{H}_{D'ABC}|_{\mathcal{S}_0}=
\nabla_{AA'}\nabla_{D'(A}\mathring{\kappa}_{BC)}|_{\mathcal{S}_0}=0.
\end{displaymath}
Now, let us define
$\mathring{\xi}_{FA'}\equiv\nabla^D_{\phantom{D}A'}\mathring{\kappa}_{DF}$.
As usual the orthogonal decomposition of this spinor is written in
terms of $\mathring{\xi}\equiv\mathring{\xi}_{AA'}\tau^{AA'}$,
$\mathring{\xi}_{AB}\equiv\tau_{(A}^{\phantom{A}C'}\mathring{\xi}_{B)C'}$.
By a computation similar to that giving equations
(\ref{xi_1})-(\ref{xi_2}) and using
(\ref{tildexi_1})-(\ref{tildexi_2}) we conclude that
$\mathring{\xi}|_{\mathcal{S}_0}=\widetilde{\xi}$,
$\mathring{\xi}_{AB}|_{\mathcal{S}_0}=\widetilde{\xi}_{AB}$. The
hypothesis that $\widetilde{\xi}$, $\widetilde{\xi}_{AB}$ fulfil
(\ref{kid_1})-(\ref{kid_2}) and a reasoning similar to that used in
the proof of proposition \ref{proposition:Killing:data} enable us to
prove
\begin{displaymath}
(\nabla_{AA'}\mathring{\xi}_{BB'} + 
\nabla_{BB'}\mathring{\xi}_{AA'})|_{\mathcal{S}_0} =0,\quad \nabla_{CC'} (\nabla_{AA'}\mathring{\xi}_{BB'} + \nabla_{BB'}\mathring{\xi}_{AA'})|_{\mathcal{S}_0}=0.
\end{displaymath}
Proposition \ref{proposition:Kspinorpropagation} now applies and
therefore we conclude that $\mathring{\kappa}_{AB}$ is a valence-2
Killing spinor in a neighbourhood of $\mathcal{S}_0$.\N

\subsection{Valence-2 Killing spinor development}
If the scalar field $\widetilde{\xi}$
defined by equation (\ref{tildexi_1}) is nonzero on $\mathcal{S}$
---that is, if the Killing initial data associated to
$\widetilde{\kappa}_{AB}$ is \emph{transversal}--- one can make use
of the notion of \emph{Killing development} introduced in \cite{BeiChr97a,BeiChr97b} to obtain a spacetime containing a valence-2 Killing spinor.  Given $(\mathcal{S},h_{ij},K_{ij})$ satisfying the Einstein (vacuum) constraints, let $\check{\mathcal{M}}=\Real \times \mathcal{S}$ and define the metric
\begin{equation}
\check{g}=\check{N}^2 \mbox{d}u^2 + \check{h}_{ij}( \mbox{d}x^i + \check{Y}^i \mbox{d}u) ( \mbox{d}x^j + \check{Y}^j \mbox{d}u), \label{Kdevelopment} 
\end{equation} 
where $\check{N}(u,x)=\mbox{Re}\widetilde{\xi}(x)$, if
$\mbox{Re}\widetilde{\xi}\neq 0$ on $\mathcal{S}$. Alternatively, if  the imaginary part of $\widetilde{\xi}$ satisfies $\mbox{Im}
\widetilde{\xi}\neq0$, then
set $\check{N}(u,x)=\mbox{Im} \widetilde{\xi}(x)$. If $\mbox{Re}\widetilde{\xi}\neq 0$, then the shift $\check{Y}^i$ is constructed by setting 
\[
\check{Y}^i(u,x)\equiv\updn{\sigma}{i}{AB}\left(\widetilde{\xi}^{AB}(x) +\widehat{\widetilde{\xi}}^{AB}(x)\right),
\]
with $\widetilde{\xi}_{AB}$ given by equation (\ref{tildexi_2}) and
$\widehat{\widetilde{\xi}}_{AB}$ its Hermitian conjugate. If
$\check{N}(u,x)=\mbox{Im} \widetilde{\xi}(x)$, then set
\[
\check{Y}^i(u,x)\equiv -\mbox{i}\updn{\sigma}{i}{AB}\left(\widetilde{\xi}^{AB}(x) -\widehat{\widetilde{\xi}}^{AB}(x)\right).
\]
If $\widetilde{\xi}$ and $\widetilde{\xi}_{AB}$ satisfy the KID
equations (\ref{kid_1}) and (\ref{kid_2}), then using well known results
about the formulation of General Relativity as a dynamical system, the
metric (\ref{Kdevelopment}) is a solution to the vacuum Einstein field
equations and $\partial_u$ is a Killing vector ---see e.g. \cite{BeiChr97a}. Now, if in addition
conditions (\ref{killingspinor_a}) and (\ref{killingspinor_b}) of
proposition \ref{proposition:Kspinor:data} hold, then $(\check{\mathcal{M}},\check{g}_{\mu\nu})$ has a valence-2
Killing spinor ---the spinor $\widetilde{\kappa}_{AB}$ is constructed
from objects with vanishing Lie derivative along the flow defined by
$(\check{N},\check{Y}^i)$. The spacetime $(\check{\mathcal{M}},\check{g}_{\mu\nu})$
is then called {\em valence-2 Killing spinor development}.

\section{Type D initial data sets}\label{section:typeD}

As discussed in the introduction, the rationale of studying conditions
on a vacuum initial data set $(\mathcal{S},h_{ij},K_{ij})$ for the
existence of Killing spinors in the development is to obtain results
enabling us to decide its Petrov type. Spacetimes containing
valence-$2$ Killing spinors are very special. Indeed, from theorem
\ref{theorem:rigidity} discussed in appendix A, these spacetimes can
only be of Petrov type N or D. The type N case can be excluded by
requiring the nonexistence of solutions to the twistor initial data
equations (\ref{twistorid_1}) and (\ref{twistorid_2}).  Hence
combining the results of the previous sections with theorem
\ref{theorem:rigidity} one obtains the following

\begin{theorem} \label{theorem:characterisationtypeD} Let
  $(\mathcal{S},h_{ij},K_{ij})$ be a suitably smooth initial data set
  for the Einstein vacuum field equations. 
Suppose that there is on $\mathcal{S}$ a valence-2 spinor $\widetilde{\kappa}_{AB}$ 
solving equations
  (\ref{killingspinor_a}) and (\ref{killingspinor_b}), and assume that 
\begin{eqnarray*}
&&\widetilde{\xi}\equiv D^{PQ}\widetilde{\kappa}_{PQ}, \\ 
&&\widetilde{\xi}_{BF}\equiv-\frac{1}{2}\widetilde{K}^{DA}_{\phantom{DA}DA}
\widetilde{\kappa}_{BF}+
\frac{3}{4}\widetilde{\kappa}^{DA}\widetilde{\Omega}_{BFDA}+\frac{3}{2}D_{(F}^{\phantom{F}D}
\widetilde{\kappa}_{B)D},
\end{eqnarray*} 
satisfy the spinorial KID equations (\ref{kid_1}) and
(\ref{kid_2}). If in addition, there is no spinor $\widetilde{\kappa}_A$ on
$\mathcal{S}$ satisfying the twistor initial data set equations
(\ref{twistorid_1}) and (\ref{twistorid_2}), then there is at least
a neighbourhood of $\mathcal{S}_0$ in the development of the initial
data where the spacetime is strictly of Petrov type D.
\end{theorem}

It is well-known that a property of the Kerr spacetime is that the
Killing vector constructed from the valence-2 Killing spinor is
degenerate, in the sense that one can always choose the phase of the
Killing spinor so that the real and imaginary parts of the (complex)
Killing vector are proportional ---see e.g. \cite{PenRin86}. This
however, does not suffice to characterise the Kerr solution ---see
e.g. \cite{FerSae07}. Possibly one may require some assumption on the
asymptotic flatness of the spacetime.  In basis of this, and using
corollary \ref{corollary:propagationreality}, one can see that
necessary ---but certainly not sufficient--- conditions for an initial
data set for the Einstein vacuum equations to be {\em Kerr initial data} are the
conditions of theorem \ref{theorem:characterisationtypeD} together with
\begin{displaymath}
\xi=\overline{\xi}, \quad \xi_{AB}=-\widehat{\xi}_{AB}.
\end{displaymath}

\section{Conclusions}
In this paper we have established conditions on a vacuum initial data
set ensuring that a neighbourhood of the initial data hypersurface is
either type N or type D. The strategy behind has been to identify the
circumstances under which the development of the initial data will be
endowed with Killing spinors. An important point of this approach is
that the characterisation is expressed in terms of
differential equations rather than in terms of algebraic conditions
on, say, the electric and magnetic parts of the Weyl tensor. Arguably,
these conditions will be hard to verify in practise. However, their
formulation in terms of a system of (possibly elliptic) overdetermined
partial differential equations could make it possible the introduction
of global arguments. The structure of at least a subset of the
conditions that has been obtained here is similar to that of the KID
equations. In \cite{Dai04c} it has been shown that it is possible to
construct certain geometric invariants for an initial data set that
indicate whether its development is static or not. It is conceivable
then, that an analogous construction for the conditions in theorem
\ref{theorem:characterisationtypeD} ---or less ambitiously in
propositions \ref{proposition:twistor:data} or
\ref{proposition:Kspinor:data}--- could be implemented rendering
geometric invariants by means of which it could be possible to decide
whether a given initial data set will give rise to Petrov type N or D
spacetime.  These ideas will be investigated elsewhere.

As it has been mentioned in several places, our results on the existence of
Killing spinors in the development of the initial data sets are local
---i.e. they only ensure the existence of the spinors in a
neighbourhood of the initial hypersurface. Global results, valid for
the maximal globally hyperbolic development can be obtained if the
spacetime and the initial data for the spinors are suitably smooth. It
is of interest whether it is possible to relax these assumptions
by using alternative arguments which do not require solving wave
equations on the whole spacetime $(\mathcal{M},g_{\mu\nu})$.

\section*{Acknowledgements}
We would like to thank KP Tod for valuable comments. We thank
CM Losert-VK for a careful reading of the manuscript.  AGP thanks the
School of Mathematical Sciences of Queen Mary College in London where
most of this work was carried out, for hospitality.  JAVK is supported
by an EPSRC Advanced Research Fellowship. AGP is supported by the
Spanish ``Ministerio de Educaci\'on y Ciencia'' under postdoctoral
grant EX-2006-0092.

\appendix

\section{A rigidity result for spacetimes with valence-$2$ Killing spinors}

The following result is used 
to give a characterisation of Petrov type D spacetimes.

\begin{proposition} \label{theorem:rigidity} 
Any vacuum Petrov type D
  spacetime $(\mathcal{M},g_{\mu\nu})$ admits a valence-$2$ Killing
  spinor. Conversely, if $\kappa_{AB}$ is a valence-2 Killing spinor
  then the spacetime is either of Petrov type D ---and $\kappa_{AB}$
  is non-degenerate, i.e. its has two different principal spinors--- or
  the spacetime is of Petrov type N ---and $\kappa_{AB}$ is
  degenerate.
\end{proposition}

\textbf{Proof.}  Assume first that $(\mathcal{M},g_{\mu\nu})$ is of
type D. Then the Weyl spinor $\Psi_{ABCD}$ has two principal spinors
$o_{A}$, $\iota_A$ in terms of which it takes the form
$$
\Psi_{ABCD}=6\psi o_{(A}o_B\iota_{C}\iota_{D)},\ o_{A}\iota^A=1.
$$
Then it is known ---see e.g. \cite{WalPen70}--- that the spinor
$\kappa_{AB}$ defined by
$$
\kappa_{AB}\equiv\psi^{-1/3}o_{(A}\iota_{B)},
$$
is a Killing spinor. Conversely, if the spinor $\kappa_{AB}$ is a
Killing spinor on a spacetime $(\mathcal{M},g_{\mu\nu})$ then we
distinguish two separate cases.

{\bf Case A:} the spinor $\kappa_{AB}$ is non-degenerate. This means
that $\kappa_{AB}=2\omega o_{(A}\iota_{B)}$, with $o_{A}\iota^A=1$.
This case was studied in \cite{Jef84} and it was shown by means of the
GHP formalism that the only possible Petrov type for $\Psi_{ABCD}$
is D.

{\bf Case B:} the spinor $\kappa_{AB}$ is degenerate. Therefore,
$\kappa_{AB}$ takes the form $\kappa_{AB}=\omega o_{A}o_B$. Let
$\iota_A$ be any spinor such that $o_A\iota^A=1$ and regard
$\{o_A,\iota_A\}$ as the spin basis used in the Newman-Penrose
formalism. Our conventions for the Newman-Penrose formalism follow
\cite{Stewart91}. Expanding the condition
$\nabla_{A'(A}\kappa_{BC)}=0$ in the Newman-Penrose spin basis and
simplifying the resulting conditions we obtain 
\begin{eqnarray*}
&& D\omega=-2\omega(\epsilon+\rho),\quad \Delta\omega=-2\gamma\omega,\\
&& \delta\omega=-2(\beta+\tau)\omega,\quad \overline{\delta}\omega=-2\alpha\omega \\
&&\sigma=0,\quad\kappa=0.
\end{eqnarray*}
We use this information in the Newman-Penrose commutation relations
which are thus reduced to
\begin{eqnarray*}
&&\hspace{-1cm}D\beta _{}^{}=-\bar{\alpha }_{}^{} \left(\epsilon _{}^{}+\rho _{}^{}\right)+\bar{\Pi }_{}^{} \left(\epsilon _{}^{}+\rho
_{}^{}\right)-\beta _{}^{} \left(\bar{\epsilon }_{}^{}+\rho _{}^{}-\bar{\rho }_{}^{}\right)+\left(\epsilon _{}^{}-\bar{\epsilon }_{}^{}+\bar{\rho
}_{}^{}\right) \tau_{}^{}-D\tau _{}^{}+\delta\epsilon _{}^{}+\delta\rho _{}^{},\\
&&\hspace{-1cm}D\alpha _{}^{}= \alpha _{}^{}
\left(-2 \epsilon _{}^{}+\bar{\epsilon }_{}^{}\right)-\left(\bar{\beta }_{}^{}-\Pi _{}^{}\right) \left(\epsilon _{}^{}+\rho _{}^{}\right)+\overline{\delta}\epsilon
_{}^{}+\overline{\delta}\rho _{}^{},\\
&&\hspace{-1cm}D\gamma _{}^{}=-\bar{\gamma }_{}^{} \left(\epsilon _{}^{}+\rho _{}^{}\right)-\gamma _{}^{} \left(2
\epsilon _{}^{}+\bar{\epsilon }_{}^{}+\rho _{}^{}\right)+\alpha _{}^{} \left(\bar{\Pi }_{}^{}+\tau _{}^{}\right)+\left(\beta _{}^{}+\tau _{}^{}\right)
\left(\Pi _{}^{}+\bar{\tau }_{}^{}\right)+\Delta\epsilon _{}^{}+\Delta\rho _{}^{},\\
&&\hspace{-1cm}\Delta\alpha _{}^{}= \bar{\beta }_{}^{}
\gamma _{}^{}+\alpha _{}^{} \left(\bar{\gamma }_{}^{}-\bar{\mu }_{}^{}\right)+\nu _{}^{} \left(\epsilon _{}^{}+\rho _{}^{}\right)-\lambda _{}^{}
\left(\beta _{}^{}+\tau _{}^{}\right)-\gamma _{}^{} \bar{\tau }_{}^{}+\overline{\delta}\gamma _{}^{},\\
&&\hspace{-1cm}\Delta\beta _{}^{}= \bar{\alpha }_{}^{}
\gamma _{}^{}-\alpha _{}^{} \bar{\lambda }_{}^{}+\beta _{}^{} \left(2 \gamma _{}^{}-\bar{\gamma }_{}^{}-\mu _{}^{}\right)+\bar{\nu }_{}^{} \left(\epsilon
_{}^{}+\rho _{}^{}\right)-\left(\bar{\gamma }_{}^{}+\mu _{}^{}\right) \tau _{}^{}-\Delta\tau _{}^{}+\delta\gamma _{}^{},\\
&&\hspace{-1cm}\delta\alpha
_{}^{}=\left(\mu _{}^{}-\bar{\mu }_{}^{}\right) \left(\epsilon _{}^{}+\rho _{}^{}\right)+\gamma _{}^{} \left(\rho _{}^{}-\bar{\rho }_{}^{}\right)+\alpha
_{}^{} \left(\bar{\alpha }_{}^{}-2 \beta _{}^{}-\tau _{}^{}\right)+\bar{\beta }_{}^{} \left(\beta _{}^{}+\tau _{}^{}\right)+\overline{\delta}\beta _{}^{}+\overline{\delta}\tau
_{}^{}.
\end{eqnarray*}
Finally we combine these conditions with the Newman-Penrose ``field
equations''. After some manipulations in the resulting set of
equations we obtain the conditions
$$
\Psi_0=\Psi_1=\Psi_2=\Psi_3=0,
$$
thus proving that $(\mathcal{M},g_{\mu\nu})$ is of Petrov type N.\N

\section{Completion of the proof of proposition \ref{proposition:Kspinorpropagation}}
\label{section:appendixB}
We fill in the details of the calculations needed to prove
proposition \ref{proposition:Killing:propagation}. First of all we
transform the identity (\ref{identiy_2}) by means of
(\ref{identity:nablas}) and insert the result into (\ref{s1})
obtaining
\begin{displaymath}
 S_{CC'DD'}=-\nabla_{D(D'}\nabla^{A}_{\phantom{A}C')}\kappa_{CA}-
\nabla_{C(D'}\nabla^{A}_{\phantom{A}C')}\kappa_{DA}.
\end{displaymath}
The covariant derivatives in this expression can be commuted using the
spinor Ricci identity with the result
\begin{displaymath}
S_{CC'DD'}=-\nabla^A_{\phantom{A}C'}\nabla_{D'(D}\kappa_{C)A}-
\nabla^A_{\phantom{A}D'}\nabla_{C'(D}\kappa_{C)A}. 
\end{displaymath}
If in this equation we use the identity $\nabla_{D'(D}\kappa_{C)A}=(H_{D'CDA}-\nabla_{AD'}\kappa_{CD})/2$ we get
\begin{displaymath}
 S_{CC'DD'}=-\frac{1}{2}\nabla^A_{\ C'}H_{D'CDA}-\nabla_{A(C'}\nabla^A_{\phantom{A}D')}\kappa_{CD}.
\end{displaymath}
Finally, if we use the identity (\ref{identity:nablas}) and the
fact that in a vacuum spacetime $\DAl_{A'B'}\kappa_{CD}=0$ we end up
with equation (\ref{covdiv_H}).


\end{document}